\documentclass[pra,showpacs,twocolumn]{revtex4-1}

\usepackage{graphicx}
\usepackage{dcolumn}
\usepackage{amsmath}
\usepackage{amssymb}
\usepackage{color}
\usepackage{comment}

\def \SiV{$\textrm{SiV}^-$~}

\newcommand{\bra}[1]{\langle #1 |}
\newcommand{\ket}[1]{| #1 \rangle}

\begin{document}

\title{Cooling phonons with phonons:\\ acoustic reservoir-engineering with silicon-vacancy centers in diamond}

\author{K. V. Kepesidis$^1$, M.-A. Lemonde$^1$, A. Norambuena$^2$, J. R. Maze$^2$, P. Rabl$^1$} 
%\email{kosmas.kepesidis@ati.ac.at}
\affiliation{$^1$Vienna Center for Quantum Science and Technology,
Atominstitut, TU Wien, 1020 Vienna, Austria}
\affiliation{$^2$Facultad de Fisica, Pontificia Universidad Catolica de Chile, Santiago 7820436, Chile}

\date{\today}

%%%%%%%%%%%%%%%%%%%%%%%%%%%%%%%%%%%%%%%%%%%%%%

\begin{abstract}
We study a setup where a single negatively-charged silicon-vacancy center in diamond is magnetically coupled to a low-frequency mechanical bending mode and via strain to the high-frequency phonon continuum of a semi-clamped diamond beam. We show that under appropriate microwave driving conditions, this setup can be used to induce a laser cooling like effect for the low-frequency mechanical vibrations, where the high-frequency longitudinal compression modes of the beam serve as an intrinsic low-temperature reservoir. We evaluate the experimental conditions under which cooling close to the quantum ground state can be achieved and describe an extended scheme for the preparation of a stationary entangled state between two mechanical modes. By relying on intrinsic properties of the mechanical beam only, this approach offers an interesting alternative for quantum manipulation schemes of mechanical systems, where otherwise efficient optomechanical interactions are not available. 
\end{abstract}

%%%%%%%%%%%%%%%%%%%%%%%%%%%%%%%%%%%%%%%%%%%%%%

\maketitle

%%%%%%%%%%%%%%%%%%%%%%%%%%%%%%%%%%%%%%%%%%%%%%%%%%%%%%%%%%%%%%%%%%%%%%%%%%%%%%%%%%%%%%%%%%%%

\section{Introduction}
Optomechanical cooling schemes are frequently used to cool individual vibrational modes of micro- and nano-mechanical  oscillators close to the quantum ground state~\cite{OMReview}. Similar to laser cooling techniques for atoms, optomechanical cooling relies on the upconvertion of low-frequency vibrational quanta into high-frequency photons, for example, by making use of the  nonlinear radiation pressure coupling between an optical cavity field and a moving mirror~\cite{WilsonRaePRL2007,MarquardtPRL2007,GenesPRA2008}.  These photons then decay into the vacuum of the electromagnetic environment, thereby removing both energy and entropy from the system. At cryogenic temperatures,  the same principle can be implemented using microwave photons~\cite{TeufelNature2011}, and instead of relying on radiation pressure, various alternative schemes using quantum dots~\cite{WilsonRaePRL2004,ZippilliPRL2009}, defect centers~\cite{RablPRB2009,KepesidisPRB2013} or superconducting two-level systems~\cite{MartinPRB2003,ZhangPRL2005,JaehneNJP2009} have been suggested to achieve equivalent cooling effects.

As first discussed in the context of trapped ions, laser cooling can be viewed as a special case of the more general concept of \emph{quantum reservoir engineering}~\cite{PoyatosPRL1996}, which refers to techniques for preparing  a quantum system in a highly non-classical stationary state by an appropriately designed dissipation mechanism~\cite{KrausPRA2008,SchirmerPRA2010,Muller2012}. A basic example of such reservoir engineering concepts is the dissipative preparation of a squeezed motional state of a trapped ion via a two-tone driving~\cite{CiracPRL1993,KienzlerScience2015}, but more elaborate schemes can be used to create, for example, highly nonclassical cat states~\cite{PoyatosPRL1996,RoyPRA2015}, stationary entangled states between separated systems~\cite{ClarkPRL2003,KrauterPRL2011,MuschikPRA2011}, or even for the preparation of non-trivial many-body states~\cite{DiehlNatPhys2008,WeimerNatPhys2010,BarreiroNature2011}. Reservoir-engineering ideas have also been discussed and implemented for macroscopic mechanical objects~\cite{RablPRB2003,KronwaldPRA2013,WollmanScience2015,WangPRL2013,TanPRA2013,TomadinPRA2012}, using again the optical or microwave radiation field as a low-entropy environment.

In this work, we describe a new approach for mechanical cooling and dissipation engineering for the low-frequency vibrations of a mechanical beam, which uses of the naturally occurring low-temperature bath provided by the high-frequency compression modes of the beam structure. Since intrinsic nonlinear interactions between mechanical modes are typically very weak, we propose here to mediate the coupling via a single electronic defect center embedded in the host lattice of the beam. Specifically, we illustrate this scheme for the example of a negatively charged silicon-vacancy (SiV$^-$) center in a vibrating diamond nanobeam, but the basic concept could be applicable for other defects and host materials as well. 

The \SiV defect is currently very actively explored as a narrow linewidth optical emitter~\cite{ClarkPRB1995,SipahigilPRL2012,NeuNJP2013,DietrichNJP2014}. However, for the present purpose, we are only interested in the electronic ground states of this center, which has both orbital and spin degrees of freedom. It exhibits an intrinsic level splitting of about 50 GHz, mainly due to spin-orbit coupling~\cite{GossPRB2007, HeppPRL2014}. This large splitting between different orbital states leads to a strong strain-induced coupling to the continuum of compression modes of the beam, which already at experimentally convenient temperatures of $T\lesssim 2.4$ K represents an effective zero-temperature reservoir. Furthermore, the low-frequency bending modes of the beam can be coupled via magnetic field gradients to the spin degrees of freedom of the \SiV defect~\cite{RablPRB2009, ArcizetNatPhys2011,KolkowitzScience2012}. By applying appropriate microwave driving fields, an efficient coupling between the two types of phonon modes can thus be engineered. 

We analyze the application of this general scheme for the ground state cooling of the fundamental bending mode, as well as for the preparation of a stationary entangled state between two different mechanical resonator modes. These schemes rely solely on the intrinsic properties of the beam and do not require optical fields or strong optomechanical couplings to microwave circuits. Thus, such phononic reservoir engineering ideas could provide a valuable alternative for mechanical systems, where an efficient integration with optical or microwave photons is not available. In view of the recent progress in the fabrication of high-quality diamond structures and mechanical beams~\cite{Ovartchaiyapong2012,BurekNanoLett2012,Tao2012}, and  demonstrations of strain-induced control of defects~\cite{MacQuarriePRL2013,Ovartchaiyapong2015,BarfussNatPhys2015,MeesalaPRApp2016,GolterPRL2016}, the proposed scheme could realistically be implemented in such systems.

\begin{figure}
  \centering
    \includegraphics[width=0.45\textwidth]{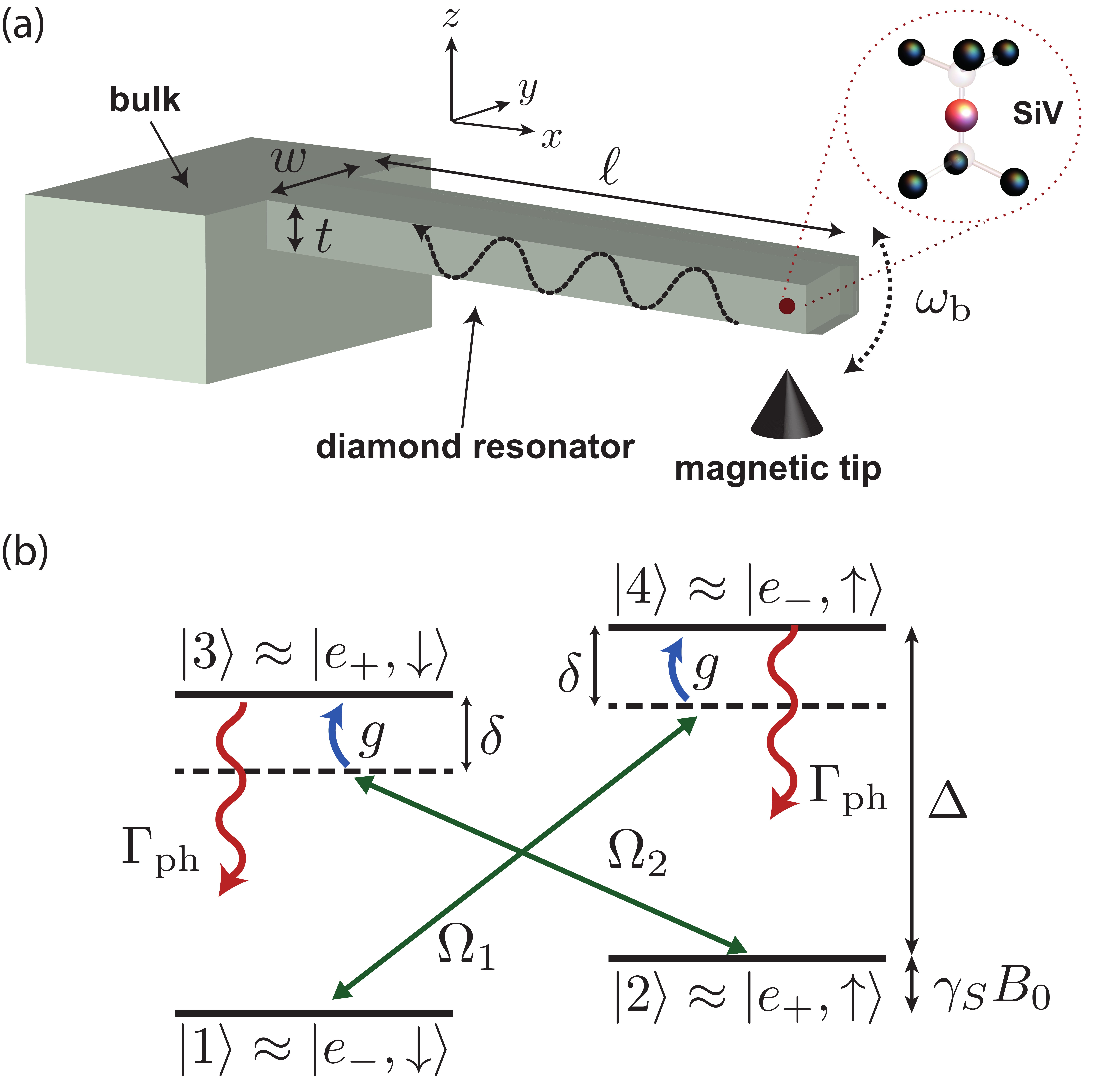}
      \caption{(a) Sketch of a single \SiV center embedded near the freely vibrating end of a diamond cantilever of length $\ell$, width $w$ and thickness $t$. In the presence of a strong magnetic field gradient produced by a nearby magnetized tip, the motion of the beam modulates the local magnetic field and results in a magnetic coupling between the electronic spin of the \SiV center and the fundamental bending mode with frequency $\omega_{\rm b}$. In addition, the orbital states of the defect are coupled via strain to a continuum of compression modes propagating along the beam (indicate by the curvy arrow). (b) Level scheme of the \SiV ground state and the relevant microwave transitions used for ground state cooling of the bending mode.}
\label{Fig:Schema}
\end{figure}

\section{Model}
We consider a setup as depicted in Fig.~\ref{Fig:Schema} (a), which shows a singly-clamped diamond nanobeam with a single \SiV center located near its freely vibrating end. In the presence of a strong magnetic field gradient, which is produced, for example, by a nearby magnetized tip, the bending motion of the beam modulates the local magnetic field and thereby couples to the electronic spin of the defect~\cite{RablPRB2009,ArcizetNatPhys2011,KolkowitzScience2012}. In addition, local lattice distortions associated with internal compression modes of the beam affect the defect's electronic structure and result in a strain coupling between these phonons and the orbital degrees of freedom of the center.  
The Hamiltonian for the whole system is given by
\begin{equation}\label{Eq:H}
\hat{H}= \hat{H}_{\rm SiV} + \hat{H}_{\rm ph} + \hat{H}_{\rm mag} + \hat{H}_{\rm strain},
\end{equation} 
where the individual terms describe the bare \SiV center, the phonon modes of the diamond beam and the magnetic and strain-induced defect-phonon couplings, respectively.

\subsection{Electronic structure of the \SiV center} \label{Sec:EStructure}
 
The negatively charged SiV color center in diamond is formed by a silicon atom and a split vacancy replacing two neighboring carbon atoms [see Fig.~\ref{Fig:Schema} (a)]. The electronic ground state of this center consists of a single unpaired hole with spin $S=1/2$, which can occupy one of the two degenerate orbital states $|e_x\rangle$ or $|e_y\rangle$~\cite{GossPRB2007, HeppPRL2014}. The spin and orbital degeneracy is lifted by the spin-orbit coupling and by the Jahn-Teller (JT) effect \cite{AbtewPRL2011, SlonczewskiPR1963}. In the presence of an external magnetic field $ \vec B$, the Hamiltonian for the electronic ground state of the \SiV center is ($\hbar=1$)~\cite{HeppPRL2014}
\begin{align} \label{Eq:HSiV}
\hat{H}^0_\textrm{SiV}  = -\lambda_\textrm{SO}\hat{L}_z \hat{S}_z + \hat{H}_\textrm{JT}  + f \gamma_L B_z \hat{L}_z +\gamma_S \vec B \cdot \vec{\hat{S}}.
\end{align}
Here $\hat{L}_z$ and $\hat{S}_z$ are the projections of the dimensionless angular momentum and spin operators $\vec{\hat{L}}$ and $\vec{\hat{S}}$ onto 
the symmetry axis of the center, which we assume to be aligned along the $z$-axis. $\lambda_\textrm{SO} > 0$ is the spin-orbit coupling while $\gamma_L$ and $\gamma_S$ are respectively the orbital and spin gyromagnetic ratios. The parameter $f\approx 0.1$ accounts for the reduced orbital Zeeman effect in the crystal lattice~\cite{HeppPRL2014, AbtewPRL2011}. 

For weak external magnetic fields, the dominant interaction in Eq.~\eqref{Eq:HSiV} is set by the spin-orbit coupling, $\lambda_{\rm SO}/2\pi\approx 45$ GHz, which splits the ground state manifold into two lower states $\lbrace |e_-,\downarrow\rangle, |e_+,\uparrow\rangle \rbrace$ and two upper states $\lbrace |e_+,\downarrow\rangle, |e_-,\uparrow\rangle \rbrace$. Here $|e_\pm\rangle=(|e_x\rangle \pm i |e_y\rangle)/\sqrt{2}$ are eigenstates of the angular momentum operator, $\hat{L}_z |e_\pm\rangle = \pm |e_\pm\rangle$.
The JT interaction $\hat{H}_{\rm JT}$ with strength $ \Upsilon<\lambda_{\rm SO}$ does not affect the spin degrees of freedom, but leads to a mixing between the orbital angular momentum states $|e_+\rangle$ and $|e_-\rangle$. 
From a combined diagonalization of the  spin-orbit and the JT interaction, we obtain a total splitting of the ground state of $\Delta=\sqrt{\lambda_{\rm SO}^2+4\Upsilon^2}\approx 2\pi \times 50$ GHz and two sets of pairwise degenerate eigenstates  $\{|1\rangle, |2 \rangle\}$ and $\{|3\rangle,|4 \rangle\}$ [cf.~Fig.~\ref{Fig:Schema} (b)]. For the purpose of this work, we can neglect the small distortions of the orbital states by the JT effect and for notational simplicity we will, in the remainder of this work, use the approximation $|1\rangle\approx |e_-,\downarrow\rangle$, $|2\rangle\approx |e_+,\uparrow\rangle$, $|3\rangle\approx |e_+,\downarrow\rangle$ and $|4\rangle\approx |e_-,\uparrow\rangle$. A more detailed discussion of the exact eigenstates and the validity of this approximation is given in Appendix \ref{App:JTint}. 

For the cooling and entangling schemes discussed below, we assume the presence of a weak static magnetic field $\vec B = B_0\vec e_z$ and additional microwave fields with frequencies $\omega_i\sim \Delta$ and amplitudes $B_i$ along $\vec e_x$.
By making a rotating wave approximation (RWA) and neglecting the reduced orbital Zeeman splitting ($\sim f\gamma_L B_z$), we obtain the following Hamiltonian for the driven \SiV center,
\begin{align} \label{Eq:Hinib}
\begin{split}
\hat{H}_\textrm{SiV} & \simeq  \left(-\Delta \hat{L}_z + \gamma_S B_0 \right)\hat{S}_z \\
&+ \left( \frac{\Omega_1}{2}|4\rangle\langle 1| e^{-i\omega_1 t} + \frac{\Omega_2}{2}|3\rangle\langle 2| e^{-i\omega_2 t} + {\rm H.c.}\right),
\end{split}
\end{align}
where $\Omega_i\sim B_i$ are the Rabi frequencies of the microwave driving fields. 
Eq.~\eqref{Eq:Hinib} is valid in the limit of weak Rabi frequencies, i.e.~$\Omega_{1,2} \ll \gamma_S B_0, \Delta$. 
The level scheme and transitions associated with $\hat{H}_{\rm SiV}$ are summarized in Fig.~\ref{Fig:Schema} (b).

\subsection{Coupling to phonon modes}
In our model, we include both magnetic and strain-induced interactions between the \SiV center and the phonon modes of the beam. Since  these two types of coupling affect the various vibrational modes differently, we divide the phonon modes into low-frequency (l.f.) and high-frequency (h.f.) modes, i.e.
\begin{equation}
\hat{H}_{\rm ph} =  \sum_l^{\rm (l.f.)} \omega_l \hat{b}^\dag_l \hat{b}_l  +  \sum_n^{\rm (h.f.)} \omega_n \hat{c}^\dag_n \hat{c}_n,
\label{Eq:Hphonon}
\end{equation} 
where the $\hat{b}_l$ and $\hat{c}_n$ are bosonic lowering operators. The first sum accounts for the lowest order vibrational modes of the beam. In particular, we are interested in the fundamental bending mode along $z$ with frequency $\omega_{\rm b}$; for a rectangular beam of dimensions $(\ell,w,t)\approx (25,0.1,0.1)\mu$m, $\omega_{\rm b} \approx \sqrt{Et^2/12\rho}(1.88/\ell)^2\approx 2\pi\times 480$ kHz with $E$ and $\rho$ being the Young's modulus and the mass-density of diamond, respectively~\cite{Cleland_Book, KleinDRM1993}. 
This frequency is too low to induce transitions between the orbital states, but the bending motion leads to a large absolute displacement of the \SiV center located at the end of the beam.  In the presence of a strong magnetic field gradient, $\partial_z B_z$, it results in a modulation of the Zeeman splitting and a magnetic spin-phonon interaction of the form~\cite{HybridOMBook}
\begin{equation}
\hat{H}_{\rm mag} =  g_{\rm m} (\hat{b}^\dag + \hat{b}) \hat{S}_z. 
\label{Eq:Hmag}
\end{equation}
Here, $g_{\rm m}= \gamma_S z_\textrm{ZPF} \partial_z B_z/\hbar$ denotes the coupling per phonon, where $z_\textrm{ZPF} =  \sqrt{\hbar/(2m\omega_{\rm b})}$ is the zero-point motion of the fundamental bending mode of the cantilever of volume $V = \ell w t$ and an effective vibrating mass of $m=\rho V/4$~\cite{Cleland_Book}. For the beam dimensions given above and achievable magnetic gradients of up to $10^7~$T/m~\cite{MaminAPL2012}, the resulting coupling strength can be as large as $g_{\rm m}/2\pi \approx 80~$kHz, and scales as $g_{\rm m}\sim \sqrt{\ell/t^2w}$ with the beam dimensions (the thickness $t$ is along the magnetic gradient). Note that while higher order bending modes would couple to the spin as well, all applications discussed below rely on resonant excitation schemes that single out a specific vibrational mode. This justifies the assumed single-mode approximation in Eq.~\eqref{Eq:Hmag} for the magnetic coupling.
 
The second sum in Eq.~\eqref{Eq:Hphonon} accounts for compression modes inside the beam with frequencies $\omega_n \sim \Delta$ of the order of the \SiV level splitting. These modes have a negligible effect on the absolute displacement of the beam, but they induce a local crystal strain which couples to the orbital states of \SiV center. 
Within the framework of linear elasticity theory, 
%and by neglecting again the distortion of the orbital states by the JT effect, 
the strain coupling is given by
%\begin{equation}
%	\hat{H}_\textrm{strain} = g_1(\hat{\gamma}_{xx} - \hat{\gamma}_{yy}) (\hat{J}_- + \hat{J}_+) - 2ig_2 \hat{\gamma}_{xy} (\hat{J}_- - \hat{J}_+).
%	\label{Eq:Hstrain1}
%\end{equation}
\begin{equation}
	\hat{H}_\textrm{strain} = g_1(\hat{\gamma}_{xx} - \hat{\gamma}_{yy}) (\hat{L}_- + \hat{L}_+) - 2ig_2 \hat{\gamma}_{xy} (\hat{L}_- - \hat{L}_+),
	\label{Eq:Hstrain1}
\end{equation}
where $\hat{L}_+ = \hat{L}_-^\dag = \ket{3}\bra{1} + \ket{2}\bra{4}$ is the orbital raising operator within the ground state
and $g_{1,2}$ are the strength of the couplings to the strain field.
From measurements of the related NV$^{-}$ defect, we expect the strain couplings $g_1 \approx g_2$ to be of the order of PHz~\cite{LeeArXiV2016}. 
The local strain fields are defined as
\begin{equation}
	\hat{\gamma}_{ij} = \frac{1}{2}\left( \frac{\partial \hat{u}_i}{\partial x_j} + \frac{\partial \hat{u}_j}{\partial x_i}\right),
\end{equation}
with $\hat{u}_{1}$ ($\hat{u}_{2}$, $\hat{u}_{3}$) representing the quantized displacement fields along $x_1 = x$ ($x_2 = y, x_3 = z$) at the position of the \SiV center. 
Eq.~\eqref{Eq:Hstrain1} assumes a set of axes as shown in Fig.~\ref{Fig:Schema} (a).
A more detailed derivation of Eq.~\eqref{Eq:Hstrain1} using group theory arguments is presented in Appendix~\ref{App:HStrain} while corrections due to the JT orbital distortion is presented in Appendix~\ref{App:JTint}.

By decomposing the local displacement field ${\vec{\hat u}}=\sum_n ( \vec u_{n} \hat c_n + \vec u_{n}^{*}c_n^\dag)$ in terms of vibrational eigenmodes with normalized mode functions $\vec u_{n}$ and bosonic operators $\hat{c}_n$, and after making a RWA, the resulting strain coupling can be written in the general form  
\begin{equation}
	\hat{H}_\textrm{strain} \simeq \sum_n^{\rm (h.f.)}  \left(g_{{\rm s}, n}\hat{c}_n \hat{J}_+  + g_{{\rm s}, n}^* \hat{c}_n^\dag \hat{J}_-\right),
	\label{Eq:Hstrain}
\end{equation}
where $\hat{J}_- = \hat{J}^\dag_+ = \ket{1}\bra{3} + \ket{2}\bra{4}$. Note that while the operators $\hat L_\pm$ induce transitions between the orbital states $\ket{e_+}$ and $\ket{e_-}$, the operators $\hat J_\pm$ induce transitions between the higher and lower energy states. For a known set of mode functions $\vec u_n$, the couplings $g_{{\rm s}, n}$ can be derived by substituting in Eq.~\eqref{Eq:Hstrain1} the corresponding strain field $\gamma^{(n)}_{ij}$ for each mode.

\subsection{Effective model}
In summary, under the validity of the different approximations outlined above, we obtain the final Hamiltonian for the \SiV center and the vibrations of the beam,
\begin{equation}\label{eq:HFull}
\begin{split}
	\hat{H}= & \hat{H}_\textrm{SiV} + \omega_{\rm b} \hat{b}^\dag \hat{b} + g_{\rm m} (\hat{b}^\dag + \hat{b}) \hat{S}_z\\
	 &+  \sum_n \omega_n \hat{c}^\dag_n \hat{c}_n + \sum_n  \left(g_{s, n}\hat{c}_n \hat{J}_+  + g_{s, n}^*\hat{c}_n^\dag \hat{J}_-\right),
\end{split} 
\end{equation}
which we will use as a starting point for the following analysis.

\section{Phonon reservoirs} 
We are primarily interested in the steady state of the bending mode, which apart from the interaction with the \SiV center already included in Eq.~\eqref{eq:HFull}, is also weakly coupled to the continuum of phonon modes in the cantilever support. 
This coupling is characterized by the damping rate $\gamma_{\rm b}=\omega_{\rm b}/Q_{\rm b}$ and the thermal equilibrium occupation number $N_{\rm b} = N_{\rm BE}(\omega_{\rm b})$, where $Q_{\rm b}$ is the mechanical quality factor of the bending mode and $N_{\rm BE}(\omega)=(e^{\hbar \omega/ k_B T } - 1 )^{-1}$ is the Bose-Einstein distribution function for a given support temperature $T$. For frequencies of $\omega_{\rm b}\sim$ MHz and temperatures $T\sim1$ K, $N_{\rm b}$ is of the order of $10^4$. In the absence of any other interaction, the bending mode will thus relax into a highly occupied thermal state. However, for realistic Q-values of $Q_{\rm b}\approx 10^5-10^6$, the thermalization with the phonon bath is very slow so that by engineering a more efficient coupling to an effective low-entropy reservoir, a cooling into a more pure quantum state can be achieved.     

In our model, we consider only the electronic ground state of the \SiV center so that no radiative decay occurs, which is usually the main ingredient for atomic and mechanical laser cooling schemes. However, the orbital states are strongly coupled to the lattice vibrations at $\sim 50$ GHz. These modes  dissipate very quickly and can thus serve as an efficient Markovian reservoir for the SiV states. The characteristic phonon-induced decay rate, $\Gamma_{\rm ph}$, for the higher-energy orbital states $|3\rangle$ and $|4\rangle$ can be estimated from Fermi's Golden rule. At temperatures $T\rightarrow0$, it is given by $\Gamma_{\rm ph}=J(\Delta)$, where 
\begin{equation}
J(\omega)= 2\pi \sum_n |g_{{\rm s},n}|^2  \delta (\omega-\omega_n)
\label{Eq:SpecFunc}
\end{equation}
is the phonon spectral density. For temperatures $T>\hbar \Delta/k_B\approx 2.4$ K, the decay rate is enhanced by the thermal occupation number $N_{\rm c} = N_{\rm BE}(\Delta)$ and reverse transitions, e.g., from state $|1\rangle$ to state $|3\rangle$, become allowed.  Experimental studies of \SiV centers in bulk at temperatures of few Kelvin observe a bare phonon induced decay rate of $\Gamma_{\rm ph}/2\pi \approx 1.6$ MHz~\cite{JahnkeNJP2015}.

\subsection{Master equation} 
Starting from the model given in Eq.~\eqref{eq:HFull}, we can use a Born-Markov approximation to eliminate the fast dynamics of the compression modes and derive an effective master equation for the density operator $\hat{\rho}$, which describes the \SiV degrees of freedom and the bending mode. We 
write the result as
\begin{equation}
\dot{\hat{\rho}} =  \left( \mathcal{L}_{\rm b}  + \mathcal{L}_{\rm SiV} + \mathcal{L}_{\rm int}    \right)\hat \rho.
\label{Eq:MEQ}
\end{equation}
The first term describes the dynamics of the bending mode,
\begin{equation}
\dot{\hat{\rho}} = -i \omega_{\rm b} [ \hat{b}^\dag\hat b,\hat \rho] +  \mathcal{L}_{\rm th} \hat \rho,
\end{equation}
where 
\begin{equation}
\label{Eq:Lth}
 \mathcal{L}_{\rm th}\hat \rho=  \frac{\gamma_{\rm b}}{2}(N_{\rm b}+1) \mathcal{D}[\hat{b}] \hat{\rho} +  \frac{\gamma_{\rm b}}{2} N_{\rm b} \mathcal{D}[\hat{b}^\dag] \hat{\rho}
 \end{equation} 
describes the coupling to the support of the cantilever. Here we have defined $\mathcal{D}[\hat{o}]\hat{\rho} =2\hat{o}\hat{\rho} \hat{o}^\dag - \hat{o}^\dag \hat{o} \hat{\rho} - \hat{\rho} \hat{o}^\dag \hat{o}$. The second term in Eq.~\eqref{Eq:MEQ} represents the bare dynamics of the \SiV center including the phonon induced decay and excitation processes, 
\begin{equation}
\mathcal{L}_{\rm SiV}\hat{\rho} =   -i [\hat{H}_{\rm SiV}, \hat{\rho} ] +  \frac{\Gamma_{\rm ph}}{2}(N_{\rm c}+1) \mathcal{D}[\hat{J}_-] \hat{\rho} +  \frac{\Gamma_{\rm ph}}{2} N_{\rm c} \mathcal{D}[\hat{J}_+] \hat{\rho}.
\end{equation}
Finally, the last term,
\begin{equation}
\mathcal{L}_{\rm int}\hat{\rho} =   -i [g_{\rm m}(\hat{b}+\hat{b}^\dag)\hat{S}_z, \hat{\rho} ],
\end{equation}
accounts for the magnetic spin-phonon interaction. 

In view of $N_{\rm c}\ll N_{\rm b}$, it is our overall goal to use the driven \SiV center to mediate an effective coupling between the bending mode and the low-entropy reservoir of compression modes. 
The extent to which this is possible depends on various system parameters. 
A particularity of the current setting is that the bending mode of interest, as well as the phononic bath composed of the continuum of the compression modes, are determined by the same beam structure.
The coherent and incoherent processes described in Eq.~\eqref{Eq:MEQ} are thus not completely independent of each other.

\subsection{Phonon spectral density} 

\begin{figure}[]
  \centering
    \includegraphics[width=0.45\textwidth]{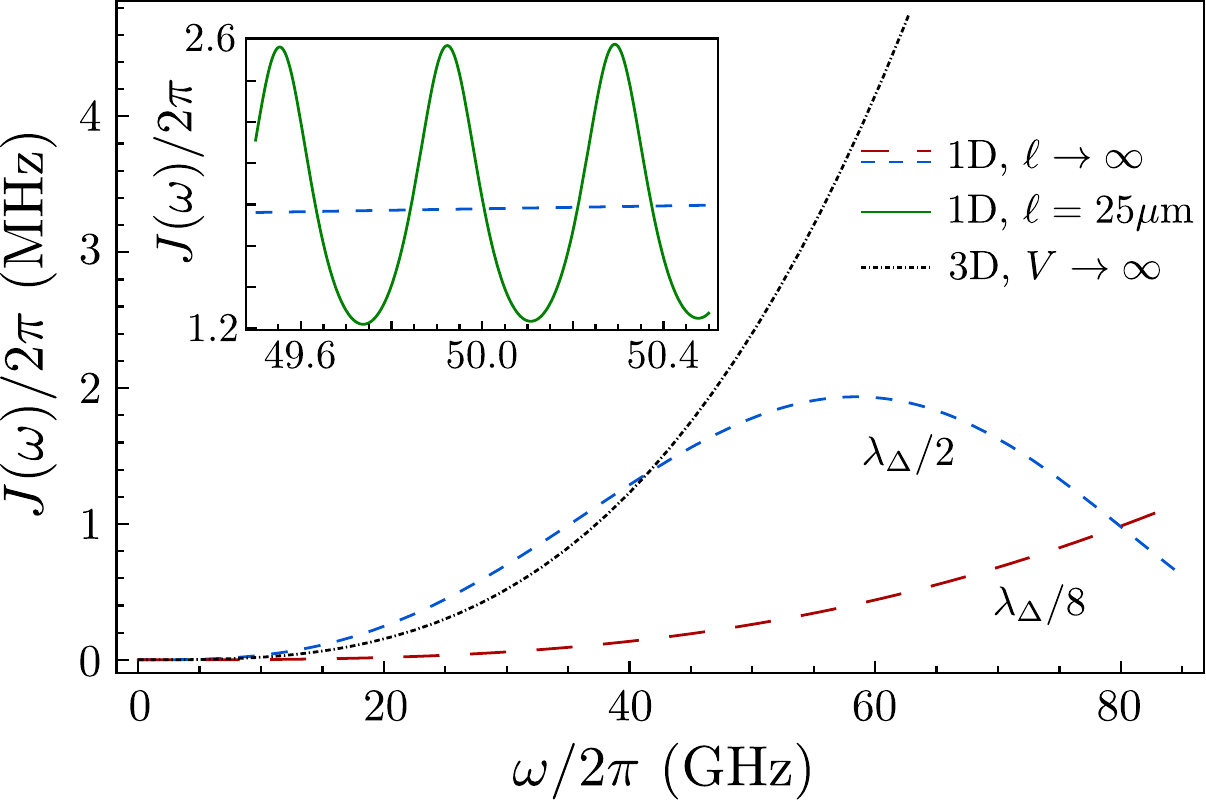}
      \caption{Spectral density of the high frequency compression modes. The blue (short-) and red (long-) dashed lines represent the spectral density in the limit of a semi-infinite 1D-cantilever along x. The density of state of the phonons is linear in frequency while the sinusoidal envelope is due to the position-dependence of the strain field. The blue and red curves are obtained for a \SiV defect positioned at $x_{\rm SiV} = \lambda_\Delta/2$ and $x_{\rm SiV} = \lambda_\Delta/8$ of the free end, respectively.
      The black dotted line represents the opposite limit of an infinite 3D-crystal. In that case, transverse modes play an important role and the phonon density of state goes as $\omega^3$. For a realistic cantilever, the 1D limit should be corrected toward the 3D-limit, due to contributions of transverse modes. In the inset, we show a finite size ($\ell = 25 \mu$m) 1D-cantilever, where the frequencies of each mode can be resolved. The finite width of each peak is due to a finite lifetime modeling losses into the bulk of the support. We chose $w=t=0.1\mu$m (1D limit), $Q_\Delta = 250$ and  $g_1 = g_2 = 2\pi \times 1$ PHz.}
      \label{Fig:PhononSD}
\end{figure}

In general, an accurate evaluation of the relevant phonon density of states requires a full numerical calculation of the individual phonon modes of a specific beam structure. However, to obtain a basic estimate of the phonon-induced decay rate and its dependence on the systems parameters, we consider here only a few limiting cases, where approximate analytic expressions can be obtained.  Furthermore, in the following analysis, we neglect the effect of the weak Poisson ratio in diamond~\cite{KleinDRM1993}.
 
As a reference, let us first consider the limit of an infinite 3D diamond crystal, which would correspond to the case where all beam dimensions are much larger than the characteristic phonon wavelength $\lambda_\Delta= \pi v/\Delta\approx 0.2$ $\mu$m, where $v = \sqrt{E/\rho}\approx 1.8\times 10^4$ m/s is the characteristic speed of sound in diamond.
In the 3D limit, the phonon modes are plane waves propagating along all directions with three possible orthogonal orientations of the displacement (polarization): one parallel (longitudinal polarization) and two perpendicular (transverse polarizations) to the wave propagation.
For identical coupling constants $g_1 = g_2\equiv g_{\rm s}$ [cf.~Eq.~\eqref{Eq:Hstrain1}], the spectral density reads (see Appendix~\ref{App:J3D})
\begin{equation}
J_{\rm 3D}(\omega)= %\frac{4}{5}\left( 2\sqrt{2} + \frac{1}{3} \right)
C\frac{\hbar g_{\rm s}^2}{\pi\rho v^5}\omega^3,
\label{Eq:J3D}
\end{equation} 
with a numerical constant $C\approx 2.5$.
By using the value $g_{\rm s}/2\pi\approx 1$ PHz, we obtain $\Gamma_{\rm ph}/2\pi \approx 1.78$ MHz, which is in good agreement with the experimental results found in Ref.~\cite{JahnkeNJP2015, LeeArXiV2016}.

The other limiting case, which is more appropriate for the considered transverse beam dimensions of $t,w\lesssim \lambda_\Delta$, is the limit of a quasi-1D beam, where the frequencies of all the transverse compression modes exceed $\Delta$. The remaining compression modes along the beam direction $x$  adopt the simple form $\vec u_n (\vec r) \propto \vec e_x  \cos(\omega_n x /v)$, with $\omega_n = \pi v n / \ell$. In that case, the strain-induced coupling constant introduced in Eq.~\eqref{Eq:Hstrain} is
\begin{equation}
	g_{{\rm s},n} = \frac{g_{\rm s}}{v} \sqrt{\frac{\hbar\omega_n}{\rho V}}\sin\left( \frac{\omega_n}{v} x_\textrm{SiV}\right),
	\label{Eq:gsn}
\end{equation}  
where $x_\textrm{SiV}$ is the position of the defect along the cantilever. 

For a completely isolated beam, the longitudinal compression modes are equally spaced with mode spacing $\delta \omega=\pi v/\ell$; for a $\ell = 25\,\mu$m long beam, $\delta\omega/2\pi\approx 370$ MHz. However, it is expected that the reflection of a compression mode at the clamping boundary is rather poor and in a realistic system, the individual resonances $\omega_n$ will be broadened by the decay $\gamma_n=\omega_n/Q_n$ into the phonon modes of the support or by other dissipation channels. 
The resulting finite phonon lifetime can be captured by approximating the phonon spectral function as a sum of individual (zero dimensional) Lorentzian peaks
\begin{equation}
	J_{\rm 0D}(\omega) \simeq \sum_n \frac{\vert g_{{\rm s},n} \vert^2}{4} \frac{\gamma_n}{(\omega -\omega_n)^2 + \gamma_n^2/4},
	\label{Eq:JD0}
\end{equation}
with $g_{{\rm s},n}$ given in Eq.~\eqref{Eq:gsn}.

In the limit $\delta\omega\lesssim \gamma_\Delta$, where $\gamma_\Delta = \Delta/Q_\Delta$ is the decay rate of the compression mode with frequency $\Delta$, the resonances of interest completely overlap and the beam can be approximated by a semi-infinite 1D beam. This limit corresponds to values of the quality factor $Q_\Delta\lesssim100$ for $\ell = 25\mu$m. The phonon spectral density then becomes
\begin{equation}
	J_{\rm 1D}(\omega) = \frac{g^2_{\rm s}}{2v^3} \frac{\hbar\omega}{\rho wt} \sin^2 \left( \frac{\omega}{v} x_\textrm{SiV}\right).
\end{equation}
We see that compared to the 3D limit, the resulting phonon induced decay rate, $\Gamma_{\rm ph} = J(\Delta)$, strongly depends on the position of the \SiV defect.
This is illustrated in Fig.~\ref{Fig:PhononSD}, where we compare the semi-infinite 1D limit for two different positions of the \SiV defect. It clearly shows the possibility to engineer smaller (larger) phonon induced decay rates by placing the \SiV close to a node (anti node) of the strain field generated by the phonon mode of frequency $\Delta$. Compared to the 3D case, the phonon induced decay in the 1D limit scales as
\begin{equation}
\frac{\Gamma_{\rm ph}^{({\rm 1D})}}{\Gamma_{\rm ph}^{({\rm 3D})}} \sim  \frac{1}{2\pi C}\left(\frac{\lambda_{\Delta}^2}{A}\right)\sin^2 \left( \pi \frac{x_\textrm{SiV}}{\lambda_\Delta} \right),
\end{equation}
where $A = wt$ is the cross section of the beam. A crossover between the 1D and the 3D limit is expected for $w\sim t \sim\lambda_{\Delta}$, i.e. when the first transverse compression mode of frequency $\Delta$ appears.

In a system where high mechanical quality factors even in the 50 GHz regime can be achieved, i.e., $Q_\Delta \gtrsim \Delta/\delta\omega $, the individual compression modes become spectrally resolved and Eq.~\eqref{Eq:JD0} applies. In this limit, the phonon spectral density can be significantly enhanced or reduced by tuning the splitting $\Delta$ in resonance or off-resonance with a compression mode. Using this tunability, the phonon induced decay could be further varied over the range
\begin{align}
	\frac{4}{\pi}\left(\frac{\Delta}{\delta \omega Q_\Delta} \right)  <  \frac{\Gamma_{\rm ph}^{\rm (0D)}}{\Gamma_{\rm ph}^{\rm (1D)}} < \frac{2}{\pi} \left(\frac{\delta \omega Q_\Delta}{\Delta} \right).
\end{align}

The dependence of $J(\omega)$ is plotted in Fig.~\ref{Fig:PhononSD} in the different limiting regimes. For the reservoir-engineering schemes discussed below, we are interested in the so-called sideband-resolved regime $\Gamma_{\rm ph}\lesssim \omega_{\rm b}$. 
This can be achieved in general for small beams $\ell\sim 10\,\mu$m, where $\omega_{\rm b}\approx 3$ MHz. In cases where the 1D limit is reached, one can further make use of the \SiV positioning and frequency tuning to reach the well-resolved sideband regime.

\section{Cooling scheme}
As a first application of the considered setup, we now analyze a cooling scheme for the low frequency bending mode. 
The basic idea is illustrated in Fig.~\ref{Fig:Schema} (b). At low temperatures, the \SiV center is predominantly in one of the lower states $|1\rangle$ or $|2\rangle$, but coupled to the excited states $|3\rangle$ or $|4\rangle$ by two microwaves fields detuned by the same amount $\delta=\omega_{1}-(\Delta+\gamma_S B_0)=\omega_2-(\Delta-\gamma_S B_0)$. In the presence of the magnetic field gradient, vibrations of the beam lead to phonon-induced processes, where the excitation of the internal state is accompanied by the absorption or emission of a motional quanta of frequency $\omega_{\rm b}$. The energy of the states $|3\rangle$ or $|4\rangle$ is then dissipated into the high frequency phonon reservoir at a rate $\Gamma_{\rm ph}$ and the cycle repeats. If the microwave fields are detuned to the red, i.e.~$\delta <0$, the phonon-absorption process dominates and the mechanical mode is cooled.   

\subsection{Reservoir engineering}

\begin{figure}[t]
  \centering
    \includegraphics[width=0.45\textwidth]{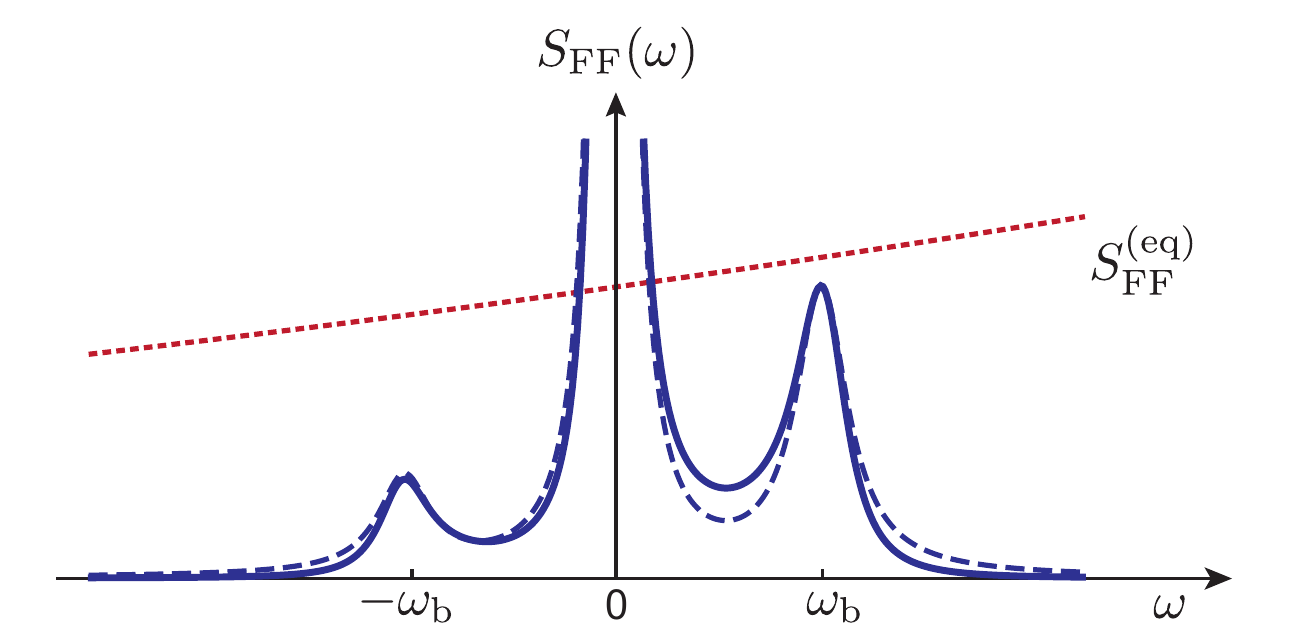}
      \caption{The force fluctuation spectrum $S_{\rm FF}(\omega)$ defined in Eq.~\eqref{Eq:SFFdef} is plotted for an \SiV center driven on the red sideband, $\delta=-\omega_{\rm b}$, and for $N_{\rm c}=0.5$ and $\Gamma_{\rm ph}/\omega_{\rm b}=0.15$.
 The dashed line corresponds to the approximate analytic result given in Eq.~\eqref{Eq:SFF}. For comparison, the red dotted line illustrates the fluctuation spectrum $S_{\rm FF}^{\rm eq}(\omega)\sim \omega\left[\coth(\hbar \omega/(2k_BT))+1\right]$ for an ohmic environment in equilibrium with a temperature $k_BT/(\hbar \omega_{\rm b})=5$.}
      \label{Fig:Spectrum}
\end{figure}

It is instructive to also look at this cooling scheme from a more general perspective by first considering the magnetic coupling [Eq.~\eqref{Eq:Hmag}] in the interaction picture with respect to the free oscillator and the \SiV Hamiltonian, 
\begin{equation}\label{eq:Hint}
\hat{H}_{\rm m}(t)= \left(\hat b e^{-i\omega_{\rm b}  t}+ \hat b^\dag e^{i\omega_{\rm b} t}\right)  \hat F(t).
\end{equation}
The operator $\hat F(t)=g_{\rm m} \hat S_z(t)$ represents a force acting on the oscillator. If we assume that the dynamics of the \SiV center is only weakly perturbed by the resonator,  i.e.,~$g_{\rm m}\sqrt{\langle b^\dag b\rangle+1} < \Gamma_{\rm ph}$~\cite{Footnote}, the fluctuations $\delta \hat F(t)= \hat F(t) - \langle\hat{F}(t)\rangle_0 $ of this force are fully characterized by the  spectrum
\begin{equation}
	S_{\rm FF}(\omega)= 2 {\rm Re}\int_0^\infty dt \,  \langle \delta \hat{F}(t)\delta \hat{F}(0) \rangle_0 e^{i\omega t}.
	\label{Eq:SFFdef}
\end{equation}
Here, $\langle \dots \rangle_0$ denotes the expectation value with respect to the  stationary state of the \SiV center in absence of the magnetic coupling. From Eq.~\eqref{eq:Hint}, we can deduce that the rates at which phonons are absorbed and emitted by the \SiV defect are proportional to the force spectrum evaluated at $+\omega_{\rm b}$ and $-\omega_{\rm b}$, respectively. For systems in thermal equilibrium, the ratio $S_{\rm FF}(-\omega)/S_{\rm FF}(\omega)=\exp(-\hbar\omega/k_BT)$ is fixed by the temperature $T$.
% which is an exact formulation of the fluctuation-dissipation theorem []. 
However, in the present case of a driven \SiV center, the system is out of thermal equilibrium and the above detailed balance equation does not hold.
The ratio $S_{\rm FF}(-\omega_{\rm b})/S_{\rm FF}(\omega_{\rm b})$ is instead used to define an effective temperature $T_{\rm eff}$ of the defect.
If the asymmetry of $S_{\rm FF}(\pm \omega_{\rm b})$ is sufficiently high, this effective temperature becomes significantly lower than the temperature of the environment, allowing the defect to act as a cold reservoir for excitations of frequency $\omega_{\rm b}$.
This is illustrated in Fig.~\ref{Fig:Spectrum}, where $S_{\rm FF}(\omega)$ is plotted for a \SiV center driven on the red sideband ($\delta =-\omega_{\rm b}$) and compared to the equilibrium spectral density of an ohmic bath.

The above heuristic arguments can be derived more rigorously from an adiabatic elimination of the \SiV degrees of freedom. The application of this technique for mechanical resonators coupled to dissipative two- or multi-level systems has been analyzed in various previous works~\cite{WilsonRaePRL2004,JaehneNJP2009,KepesidisPRB2013,CiracPRA1992,RablPRB2010}.
As a result, we obtain an effective master equation for the reduced density operator of the bending mode, $\hat \rho_{\rm b}$,
\begin{equation}
\dot{\hat{\rho}}_{\rm b} = \left( \mathcal{L}_{\rm th}+  \mathcal{L}_{\rm eff}\right)  \hat{\rho}_{\rm b}.
\label{Eq:RhoEff}
\end{equation}
The second term describes the effect of the engineered effective reservoir
\begin{equation}
 \mathcal{L}_{\rm eff}\hat \rho =   \frac{\Gamma_{\rm eff}}{2}(N_{\rm eff}+1) \mathcal{D}[\hat{b}] \hat{\rho} +   \frac{\Gamma_{\rm eff}}{2}N_{\rm eff}\mathcal{D}[\hat{b}^\dag] \hat{\rho},
\end{equation}
where $\Gamma_{\rm eff} = S_{\rm FF}(\omega_{\rm b}) - S_{\rm FF}(-\omega_{\rm b})$ and $N_{\rm eff}= S_{\rm FF}(-\omega_{\rm b})/\Gamma_{\rm eff}$. From Eq.~\eqref{Eq:RhoEff} we can derive the steady-state occupation number of the fundamental bending mode
\begin{equation}
	\bar{n}_{\rm b} = \frac{\gamma_{\rm b} N_{\rm b} + N_{\rm eff}\Gamma_{\rm eff}}{\gamma_{\rm b} + \Gamma_{\rm eff}} \simeq \frac{\gamma_{\rm b} N_{\rm b}}{\Gamma_{\rm eff}}+ N_{\rm eff},
	\label{Eq:nfinal}
\end{equation}
where $\gamma_{\rm b} N_{\rm b} \simeq k_BT/(\hbar Q_{\rm b})$ is the rate at which the mechanical system in the ground state would absorb a single phonon from the environment. The last equality is valid in the limit $\Gamma_{\rm eff} \gg \gamma_{\rm b}$.

\subsection{Ground state cooling}
The force spectrum $S_{\rm FF}(\omega)$ can be evaluated numerically for all parameters using the quantum regression theorem. To obtain more insights, we restrict the following discussion to a fully symmetric situation $\Omega_1=\Omega_2=\Omega$ under weak driving conditions $\Omega \ll  |\Gamma_{\rm ph} + i \delta|$. In this limit, the spectrum is approximately given by
\begin{equation}
\begin{split}
S_{\rm FF}(\omega) =  \epsilon & g_{\rm m}^2\Gamma_{\rm ph} \left\lbrace\frac{N_{\rm c}+1}{(\delta +\omega)^2 + [(2N_{\rm c}+1)\Gamma_{\rm ph}/2]^2} \right. \\
& + \frac{N_{\rm c}}{(\delta -\omega)^2 + [(2N_{\rm c}+1)\Gamma_{\rm ph}/2]^2} \\
& + \left. \frac{2N_{\rm c}+1}{\omega^2 + [\epsilon(2N_{\rm c}+1)\Gamma_{\rm ph}/2]^2} \right\rbrace,
\end{split}\label{Eq:SFF}
\end{equation}
where we have defined the probability to excite the higher-energy states via the drives
\begin{equation}
	\epsilon = \frac{\Omega^2/4}{\delta^2+[(1 + 2N_{\rm c})\Gamma_{\rm ph}/2]^2}.
\end{equation}
An example of the spectrum is illustrated in Fig.~\ref{Fig:Spectrum} and exhibits three peaks at frequencies $\omega=0, \pm \delta$. For red-detuned driving fields, the first term in Eq.~\eqref{Eq:SFF} corresponds to the process where an excitation of the bending motion is absorbed, while the second term corresponds to the Stokes process where a phonon is emitted. Note that for a reservoir at finite temperature, the ratio between the heating and the cooling process scales as $N_{\rm c}/ (N_{\rm c}+1)$. The peak at $\omega=0$ arises from random spin-flip processes which are induced by the driving fields even in the absence of the phonon mode. These random spin flips occur at a rate $\Gamma_{\rm sf}\simeq \epsilon (2N_{\rm c}+1)\Gamma_{\rm ph}$ and create a fluctuating force symmetric in frequency. Therefore, they only contribute to $N_{\rm eff}$ and not to the cooling rate $\Gamma_{\rm eff}$.  

From this expression and in view of Eq.~\eqref{Eq:nfinal}, we identify three basic conditions for achieving ground state cooling, i.e.~$\bar n_{\rm b}<1$. 
First, the system must be in the well-resolved sideband regime (i) $\Gamma_{\rm ph}(2N_{\rm c}+1)\lesssim \omega_{\rm b}$. Provided that $S_{\rm FF}(\omega)$ [cf.~Eq.~\eqref{Eq:SFF}] is a sum of three Lorentzian peaks centered at $\omega = 0$ and $\omega = \pm \delta$ with width $\sim \Gamma_{\rm ph}(2N_{\rm c}+1)$,
this condition is necessary to have a large asymmetry between the positive-frequency and negative-frequency domain of the force spectrum in the ideal case $\delta =  -\omega_{\rm b}$. As a consequence, $\Gamma_{\rm eff}$ is enhanced while the minimal occupancy $N_{\rm eff}$ is decreased.
If this condition is satisfied, the remaining asymmetry is determined by $N_{\rm c}$, which sets the limit $\bar n_{\rm b}\geq N_{\rm c}$. 
In terms of the effective temperature, this limit means that the cooling ratio can reach
\begin{equation}
\frac{T_{\rm eff}}{T} = \frac{\omega_{\rm b}}{\Delta} \approx 10^{-5}-10^{-4}.
\end{equation}
The high-frequency phonon reservoir must then also be close to the ground state, i.e.~(ii) $N_{\rm c}<1$.
Finally, the engineered cooling rate, 
\begin{align}
\Gamma_{\rm eff} & = \epsilon g_{\rm m}^2\Gamma_{\rm ph}\left\lbrace \frac{1}{(\delta +\omega_{\rm b})^2 + [(1 +2N_{\rm c})\Gamma_{\rm ph}/2]^2} \right. \nonumber \\
 & \qquad\qquad \left. - \frac{1}{(\delta
   -\omega_{\rm b})^2+[(1 +2 N_{\rm c})\Gamma_{\rm ph}/2]^2} \right\rbrace,
   \label{Eq:Gammaeff}
\end{align}
must overcome the rethermalization rate of the natural bath (iii) $\Gamma_{\rm eff} >\gamma_{\rm b} N_{\rm b}$. For a phononic bath in its ground state ($N_{\rm c} < 1$) coupled to a \SiV center driven on a very well-resolved red sideband ($\delta = -\omega_{\rm b}$, $\Gamma_{\rm ph} \ll \omega_{\rm b}$), the effective damping rate is maximized to $\Gamma_{\rm eff} = g_{\rm m}^2\Omega^2/(\Gamma_{\rm ph} \omega_{\rm b}^2)$.  Note that this result is derived under the assumption of weak coupling and the cooling rate is always bounded by $\Gamma_{\rm eff}\lesssim g_{\rm m}$.
%
%while the minimal occupancy reaches the minimized value $N_{\rm eff} = \Gamma_{\rm ph}^2 / 16\omega_{\rm b}^2$.

\begin{figure}[]
  \centering
    \includegraphics[width=0.45\textwidth]{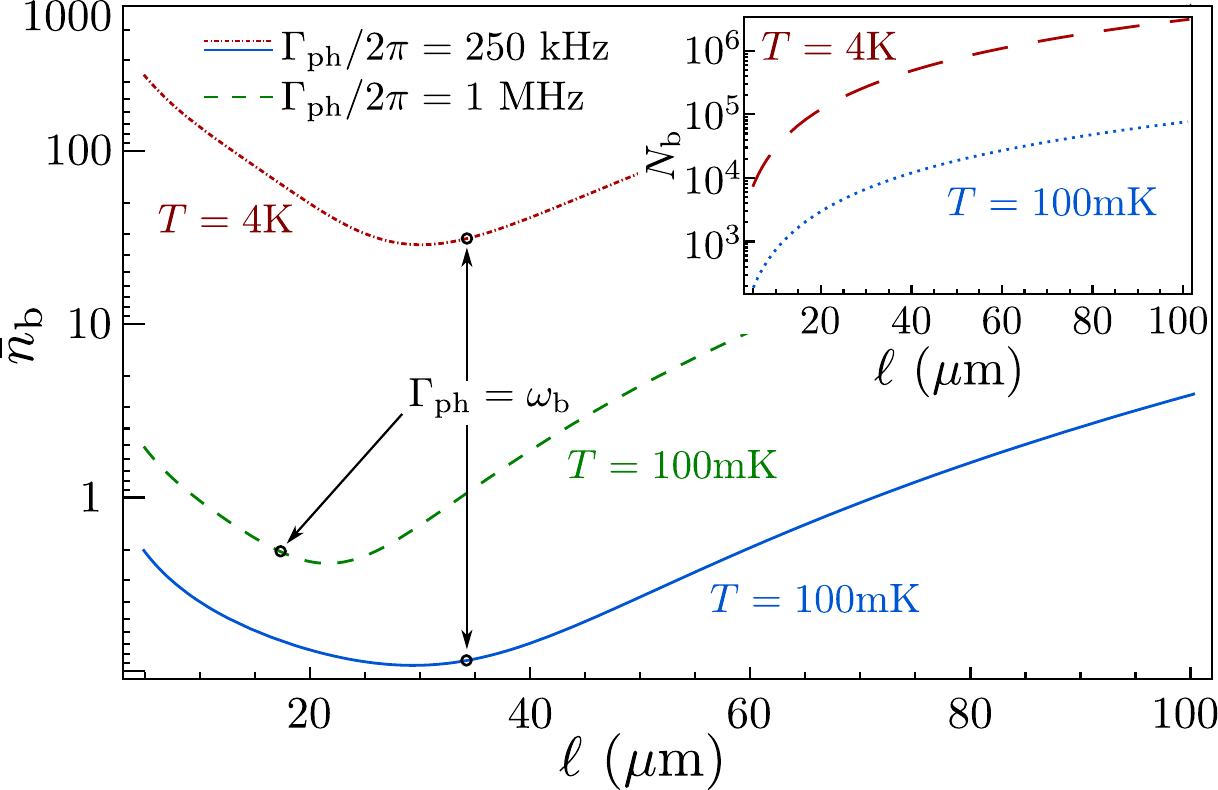}
      \caption{Final occupation of the bending mode $\bar{n}_{\rm b}$ as function of the beam length $\ell$ for different values of $\Gamma_{\rm ph}$. We considered a diamond cantilever at $T=4$K (dotted-dashed red line), which corresponds to an occupation of the compression mode of frequency $\Delta/2\pi = 50$ GHz of $N_{\rm c} \approx 1.2$.  The blue solid line shows the results for $\Gamma_{\rm ph}/2\pi = 250$ kHz and a lower temperature $T=100$mK ($N_{\rm c} \simeq 0$). In the inset, we plot the initial occupation ($g_{\rm m} = 0$) of the bending mode. It shows that even for initial occupancy $N_{\rm b} \sim 10^4$, it is possible to reach the ground state ($\bar{n}_{\rm b} < 1$) using experimentally accessible parameters. These results are obtained using a driving strength $\epsilon = 0.2$ [cf.~Eq.~\eqref{Eq:Gammaeff}] and the full expression of $S_{\rm FF}(\omega)$ that includes the drive at every order; we use the corresponding optimal detuning $\delta \approx -\omega_{\rm b}/\sqrt{1+4\epsilon}$. The remaining parameters are the bending mode quality factor $Q_{\rm b} = 10^6$, the transverse beam dimension $w = t = 0.1$ $\mu$m and a magnetic field gradient of $10^7$ T/m.}
      \label{Fig:nFinal}
\end{figure}

While condition (ii) is solely determined by the temperature of the support, the competition between the coupling strength $g_{\rm m} \propto \sqrt{\ell}$ and the mechanical frequency $\omega_{\rm b} \propto 1/\ell^2$ may prevent one from fulfilling (i) and (iii) at the same time. This is illustrated in  Fig.~\ref{Fig:nFinal}, where we show the final occupancy of the fundamental bending mode $\bar n_{\rm b}$ [cf.~Eq.~\eqref{Eq:nfinal}] as a function of the beam length $\ell$ for different values of $\Gamma_{\rm ph}$ and $T$. We see that already for temperatures $T\lesssim 1$ K and a moderate suppression of the phonon spectral density below the 3D limit, ground state cooling becomes feasible.

\section{Steady-state entanglement}

As a second application, we now discuss an extension of the previous scheme for the dissipative preparation of an entangled two-mode squeezed (TMS) state $|\psi_{\text{TMS}} \rangle$, which is shared between two low-frequency mechanical modes $\hat{a}$ and $\hat{b}$. 
A pure TMS state is defined by the conditions
\begin{equation}\begin{split}\label{Eq:TMS}
& \hat{A}|\psi_{\text{TMS}}\rangle =( \mu \hat{a} + \nu \hat{b}^{\dag}) |\psi_{\text{TMS}}\rangle=0,\\
& \hat{B}|\psi_{\text{TMS}}\rangle  =( \mu \hat{b} + \nu \hat{a}^{\dag}) |\psi_{\text{TMS}}\rangle=0,
\end{split}\end{equation}
where $\mu$ and $\nu$ are complex parameters satisfying $|\mu|^2 - |\nu|^2 = 1$. This implies that these parameters can be written as $\mu = \cosh(r)$ and $\nu = e^{i\theta}Ê\sinh(r)$, where $r$ is the squeezing parameter.  The TMS state belongs to the family of Gaussian states for which entanglement can be verified from a violation of the separability criteria~\cite{DuanPRL2000, SimonPRL2000}
%\MALcomment{This is a particular expression of the PPT criterion, which is sensitive to the phase of $\langle\hat{a}\hat{b}\rangle$.}
\begin{equation}\label{Eq:Separability}
\xi = \frac{1}{2} \left[  \text{Var}(\hat{x}_a + \hat{x}_b) + \text{Var}(\hat{p}_a - \hat{p}_b)  \right] \geq 1,
\end{equation}
where $\hat{x}_a = (\hat{a}^{\dagger} + \hat{a})/\sqrt{2}$ and $\hat{p}_a = i(\hat{a}^{\dagger} - \hat{a})/\sqrt{2}$ are the normalized position and momentum operators ($a \rightarrow b$ for the $\hat{b}$ mode). 
For a pure TMS state and $\theta=0$, the amount of entanglement increases exponentially with the squeezing parameter, i.e.~$\xi = e^{-2r}$.
%For a pure two-mode squeezed state $\xi = e^{-2r}$, meaning that an appreciable amount of entanglement can already be achieved for $r\sim 1$. 

For mechanical oscillators prepared in a pure TMS state, the thermal noise coming from their coupling to the environment rapidly degrades such a fragile entangled state.
It is thus intriguing to consider engineered processes, where the TMS state emerges as a steady state of a purely dissipative dynamics. 
From the dark state conditions \eqref{Eq:TMS}, one can readily see that $|\psi_{\text{TMS}} \rangle$ is the stationary state of a master equation of the form 
\begin{equation} 
\begin{split}
\mathcal{L}_{\rm sq} \hat{\rho} & =   \frac{\Gamma_{\rm sq}}{2}  \left( 2 \hat{A} \hat{\rho} \hat{A}^{\dagger} - \hat{A}^{\dagger} \hat{A} \hat{\rho} - \hat{\rho} \hat{A}^{\dagger} \hat{A} \right)\\
& + \frac{\Gamma_{\rm sq}}{2}  \left( 2 \hat{B} \hat{\rho} \hat{B}^{\dagger} - \hat{B}^{\dagger} \hat{B} \hat{\rho} - \hat{\rho} \hat{B}^{\dagger} \hat{B} \right).
\end{split}
\label{Eq:Lsq}
\end{equation}
The implementation of such a dissipation process has previously been discussed and experimentally implemented in the context of two separated spin ensembles coupled to a common optical channel~\cite{KrauterPRL2011,MuschikPRA2011} and closely related schemes have been proposed for optomechanical systems~\cite{WangPRL2013,TanPRA2013,HartmannPRL2008}.  In the following, we show how such an effective dissipation can be realized in the present setup and evaluate its robustness with respect to finite temperature effects and deviations from the ideal side-band resolved limit.

\begin{figure}\label{squeezingFig}
  \centering
    \includegraphics[width=0.48\textwidth]{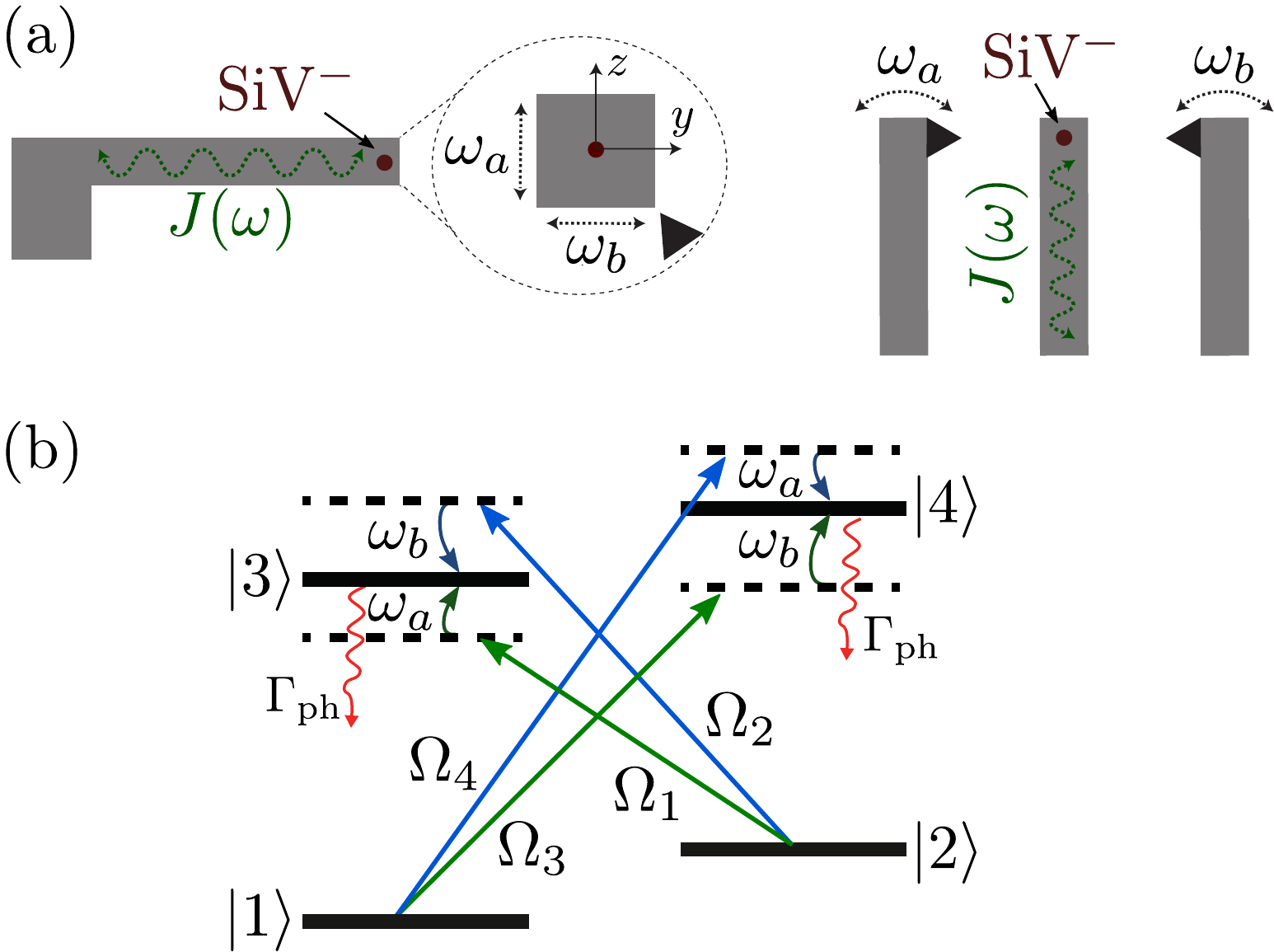}
      \caption{Setup for the dissipative preparation of a mechanical two-mode squeezed state. (a) Two possible configurations, where two vibrational modes, representing either orthogonal vibration modes of the same beam (left) or the fundamental bending modes of two different beams (right), are coupled magnetically to the same \SiV center.
      (b) Energy level diagram of the \SiV defect. The green and blue arrows indicate the four driving fields which are detuned from resonance by $\pm\omega_a$ and $\pm\omega_b$. See text for more details.
      %The damping rate $\Gamma_{\rm ph}$ comes from the coupling to the high frequency compression modes.
      }\label{Fig:Squeezing}
\end{figure}

\subsection{Driving scheme}

We consider two low-frequency mechanical modes $\hat{a}$ and $\hat{b}$ with two different frequencies $\omega_a$ and $\omega_b$, respectively. As shown in Fig.~\ref{Fig:Squeezing} (a), these two modes could be two bending modes of the same beam, or the fundamental modes of two independent beams coupled magnetically to the same \SiV center. In contrast to the cooling scheme, engineering the jump operators $\hat{A}$ and $\hat{B}$ requires driving the transitions $|1\rangle \rightarrow |4\rangle$ and $|2\rangle\rightarrow |3\rangle$ by two near-resonant microwave fields. As indicated in Fig.~\ref{Fig:Squeezing} (b), 
those drives are detuned by $-\omega_a$ and $+\omega_b$ in one half of the cycle (e.g. transitions $2\rightarrow3$) where 
a phonon is added to the mode $a$ and removed from the mode $b$, and by $+\omega_a$ and $-\omega_b$ in the other half (e.g. transitions $1\rightarrow4$) where the opposite processes take place.

The total Hamiltonian for this system is then given by
\begin{equation}
\hat{H}_{\rm sq} = \hat{H}_{\rm res} + \hat{H}_{\text{SiV}} + \hat{H}_{\rm m},
\end{equation}
where $\hat{H}_{\rm res} = \omega_a \hat{a}^{\dag} \hat{a} + \omega_b \hat{b}^{\dag} \hat{b}$ is the Hamiltonian of the two mechanical resonator modes and
\begin{align}\label{eq:HSiV4drive}
\begin{split}
\hat{H}_{\text{SiV}} & = \sum_{j=1,4} E_j\vert j \rangle \langle j \vert \\
& +\left( \frac{\Omega_1}{2} e^{-i\omega_1 t}  - \frac{\Omega_2}{2} e^{-i\omega_2 t}\right) |3\rangle \langle 2| + {\rm H.c.}\\
& -\left( \frac{\Omega_3}{2} e^{-i\omega_3 t}  - \frac{\Omega_4}{2} e^{-i\omega_4 t}\right) |4\rangle \langle 1| + {\rm H.c.}
\end{split}
\end{align}
is the Hamiltonian for the internal states, where $E_1=0$, $E_2=\gamma_SB_0$, $E_3=\Delta$ and $E_4= \Delta+\gamma_SB_0$. Note that in Eq.~\eqref{eq:HSiV4drive} all Rabi frequencies $\Omega_i$ are assumed to be real and positive and the minus signs have been chosen to reproduce the correct effective interaction in the analysis below. Finally,
\begin{equation}
\hat{H}_{\rm m} = \left[ g_a(\hat{a} + \hat{a}^{\dag})  + g_b (\hat{b} + \hat{b}^{\dag} Ê)\right] \hat{S}_z
\end{equation}
is the Zeeman coupling as introduced in Eq.~\eqref{Eq:Hmag}. 

\subsection{Sideband-resolved regime} \label{Sec:TMS_SBR}
For the implementation of the TMS state master equation~\eqref{Eq:Lsq}, it is instructive to first follow a simplified approach that is valid in the well-resolved sideband regime. 
In this regime, only the resonant processes play an important role and, as we show, are responsible for the mechanical entanglement.
In order to identify those dominant processes, we first perform a polaron transformation $\hat{H}\rightarrow \hat{U}\hat{H}\hat{U}^\dag$~\cite{WilsonRaePRL2004}, where 
\begin{equation}
\hat{U} = \exp \left\lbrace\sum_{\eta=a,b} \frac{g_\eta}{\omega_\eta}Ê(\hat{\eta}^\dag - \hat{\eta})\hat{S}_z \right\rbrace.
\end{equation}
%\begin{equation}
%U=e^{iS}, \qquad S= i \sum_{\eta=a,b} \frac{g_\eta}{\omega_\eta}Ê(\eta^\dag -\eta)S_z.
%\end{equation}
This transformation eliminates the coupling $\hat{H}_{\rm m}$, 
but generates an infinite series of phonon sidebands in the driving processes, i.e.~
%but transforms the transition operators like
\begin{align}
\begin{split}
 |3\rangle\langle 2| &\rightarrow   |3\rangle\langle 2| e^{-\sum_{\eta=a,b} \frac{g_\eta}{\omega_\eta}Ê(\hat{\eta}^\dag -\hat{\eta})}, \\
 |4\rangle\langle 1| &\rightarrow   |4\rangle\langle 1| e^{\sum_{\eta=a,b} \frac{g_\eta}{\omega_\eta}Ê(\hat{\eta}^\dag -\hat{\eta})}.
 \end{split}
\end{align}   
For $g_\eta/\omega_\eta\ll 1$, we can expand the exponential to first order and thus neglect the contribution of higher-order phonon processes. By going into the interaction picture with respect to $\hat{H}_{\rm res} + \hat{H}_{\text{SiV}}$ and making a RWA, only the resonant processes remain so that $\hat{H}_{\rm sq}$ adopts the form
\begin{equation}
\hat{H}_{\rm sq} \simeq \frac{\tilde{g}}{2}(\hat{A}  |3\rangle \langle 2| + \hat{B}  |4\rangle \langle 1|) + {\rm H.c.}
\end{equation} 
In the particular case of a symmetric driving cycle (cf.~Fig.~\ref{Fig:Squeezing}), where the cooling (heating) processes for both mechanical modes occur at the same rates, i.e.~$\Omega_1 g_a / \omega_a = \Omega_3 g_b / \omega_b$ ($\Omega_2 g_b / \omega_b = \Omega_4 g_a/ \omega_a$), the new jump operators $\hat{A}$ and $\hat{B}$ are defined as in Eq.~\eqref{Eq:TMS} with
\begin{equation}
\begin{split}
	\tilde{g}\mu = \frac{\Omega_1 g_a}{\omega_a} = \frac{\Omega_3 g_b}{\omega_b}, \qquad
	\tilde{g}\nu = \frac{\Omega_4 g_a}{\omega_a} = \frac{\Omega_2 g_b}{\omega_b}.
\end{split}
\end{equation}
The condition $|\mu|^2 - |\nu|^2 = 1$, which imposes  the proper commutation relations for $\hat{A}$ and $\hat{B}$, is fulfilled when the coupling constant $\tilde{g}$, given by
\begin{equation}
	\tilde{g} = \frac{g_a}{\omega_a}\sqrt{\Omega_1^2-\Omega_4^2} = \frac{g_b}{\omega_b}\sqrt{\Omega_3^2-\Omega_2^2},
\end{equation}
remains real. One can show that for drive strengths that lead to an imaginary coupling constant $\tilde{g}$, the coherent dynamics described by $\hat{H}_{\rm sq}$ becomes unstable. Physically, it corresponds to heating processes ($\sim \Omega_2, \Omega_4$) that exceed the rates at which the drives cool down the mechanical modes ($\sim \Omega_1, \Omega_3$), leading to a parametric instability of the mechanical system.
The squeezing parameter is also determined by the ratio between the driving strengths for the two mechanical modes:
\begin{equation}
	\tanh(r) = \frac{\Omega_2}{\Omega_3} = \frac{\Omega_4}{\Omega_1}.
\end{equation}
As expected, the squeezing parameter diverges when the heating and cooling rates become equal.

As a final step, we can adiabatically eliminate the internal degrees of freedom of the SiV$^-$, given that $\Gamma_{\rm ph} \gg g_a, g_b$, and obtain the master equation 
\begin{equation}\label{Eq:MEQsqueezing}
\dot{\hat{\rho}} = \left(\mathcal{L}_{\rm th}  + \mathcal{L}_{\rm sq}\right)  \hat{\rho}.
\end{equation}
The first term describes the coupling of the mode $\hat{a}$ and $\hat{b}$ to their respective thermal environment [cf.~Eq.~\eqref{Eq:Lth}] with damping rates $\gamma_{a/b}$ and thermal occupation numbers $N_{a/b}$. The second term describes the engineered dissipative processes that lead to entanglement, and in the limit of $N_c = 0$, it reduces to Eq.~\eqref{Eq:Lsq} with the corresponding rate $\Gamma_{\rm sq} = \tilde g^2/(2\Gamma_{\rm ph})$.  Note that in the sideband-resolved and weak driving limit this rate scales as $\Gamma_{\rm sq}\simeq\Gamma_{\rm eff}/(2\cosh(r) )$ compared to the optimized single mode cooling rate, assuming $g_a \approx g_b$ and $\omega_a \approx \omega_b$. 
%\MALcomment{Maybe add a note on neglecting the polaron transformation in the dissipation terms $\mathcal{L}_{\rm th}$.}

\begin{figure}[]
  \centering
    \includegraphics[width=0.48\textwidth]{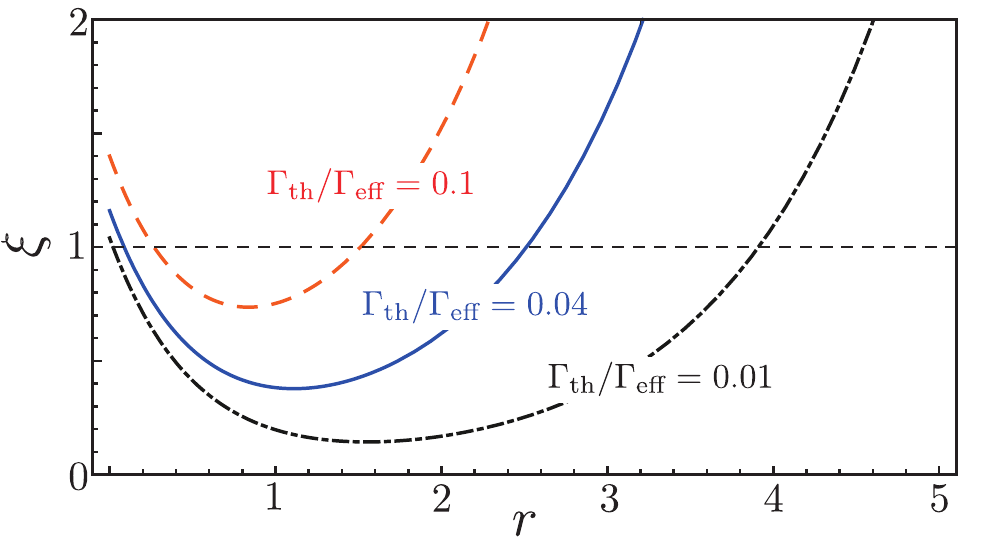}
      \caption{The separability parameter $\xi$ is plotted as a function of the squeezing parameter $r$ for three values of the ratio $\Gamma_{\text{th}} / \Gamma_{\text{eff}}$, in the limit $\gamma_{a,b}\ll \Gamma_{\rm sq}$ and $\gamma_aN_a= \gamma_b N_b=\Gamma_{\rm th}$. Vertical dotted line indicates the threshold bellow which the two bending modes are entangled, as given by Eq.~\eqref{Eq:Separability}. For this plot ideal sideband resolved conditions and $N_{\rm c}=0$ have been assumed.}\label{SeparabilityFig}
\end{figure}

As a consequence, the amount of steady-state entanglement between the two mechanical modes is a results of the competition between the squeezing rate $\Gamma_{\rm sq}$ and the rates at which the thermal noise perturbs the system, $(N_b + 1)\gamma_b$, $(N_a + 1)\gamma_a$. To make this statement more precise, one can explicitly solve Eq.~\eqref{Eq:MEQsqueezing} for the steady-state ($\dot{\hat{\rho}} = 0$), resulting in:
%\begin{equation}
%\begin{split}
%\xi =& \frac{ \gamma_a (N_a + \tfrac{1}{2}) + \gamma_b (N_b + \tfrac{1}{2}) + \Gamma_{\rm sq} e^{-2r} }{ \tfrac{\gamma_a + \gamma_b}{2} +  \Gamma_{\rm sq} }.
%\end{split}
%\end{equation}
\begin{equation}
\begin{split}
\xi =& \frac{\gamma_a N_a + \Gamma_{\text{sq}} \nu^2 }{\gamma_a + \Gamma_{\text{sq}} } +\frac{\gamma_b N_b + \Gamma_{\text{sq}} \nu^2 }{\gamma_b + \Gamma_{\text{sq}} } - \frac{2 \Gamma_{\text{sq}} \mu \nu}{ \frac{\gamma_a + \gamma_b}{2} + \Gamma_{\text{sq}}} + 1 .
\end{split}
\end{equation}
A pure TMS state can only be achieved in the limit where $\Gamma_{\rm sq}$ greatly exceeds the thermal noise rate. On the other hand, the larger is $\Gamma_{\rm sq} \sim (\Omega_1^2 - \Omega_4^2)/\Gamma_{\rm ph}$, the smaller is the final amount of squeezing $\tanh r \sim \Omega_4/\Omega_1$. By assuming $\gamma_{a,b}\ll \Gamma_{\rm sq}$ and  $\gamma_aN_a= \gamma_b N_b=\Gamma_{\rm th}$, the separability criterion can be approximated by
\begin{equation} \label{Eq:xiApprox}
\begin{split}
\xi \approx  \frac{4\Gamma_{\rm th}}{\Gamma_{\rm eff}} \cosh(r) + e^{-2r},
\end{split}
\end{equation}
where again $\Gamma_{\rm eff}=g^2_b\Omega_3^2/(\omega_b^2\Gamma_{\rm ph)}$ is the single mode cooling rate in the well-resolved sideband limit. The dependence of $\xi$ on $r$ is plotted in Fig.~\ref{SeparabilityFig} for different values of $\Gamma_{\rm th}/\Gamma_{\rm eff}$.

%The maximal amount of steady-state entanglement that can be achieved in a pure TMS state must then respect
%\begin{equation}
%	\cosh^2r \ll \max \left[ \frac{g_a^2 \Omega_1}{\omega_a^2\gamma_{\rm ph}\gamma_a(N_a + 1)}, \frac{g_b^2 \Omega_3}{\omega_b^2\gamma_{\rm ph}\gamma_b(N_b + 1)} \right].
%\end{equation}
%In Fig.~(), we show that...

\subsection{Two-mode squeezing by a finite-temperature reservoir} 
For the derivation above we have assumed $N_{\rm c}=0$ and the sideband-resolved regime, such that the temperature of the engineered reservoir is zero. We now want to evaluate how sensitive the entanglement is with respect to small deviations from these conditions. 
In principle, one can use a similar approach as in Sec.~\ref{Sec:TMS_SBR} to derive an effective master equation for the two mechanical modes.
All the rates are then determined by the fluctuation spectrum $S_{\rm FF}(\omega)$ (including the four drives) evaluated at the relevant frequencies $\omega=\pm \omega_a, \pm \omega_b$. However, outside the sideband-resolved regime the result would be quite involved and not very illuminating. 
Instead, we will here analyze an approximated master equation, which nevertheless captures the most essential effects of a finite effective temperature on the entanglement generation and can thus be adopted to other systems as well.

Let us first point out that when generalized to a finite $N_{\rm c}$, the above approach introduces for every dissipation processes associated with the jump operator $\hat A$ ($\hat B$), a reverse process with operator $\hat A^\dag$ ($\hat B^\dag$). The rates of the original and the reverse processes scale like $(N_{\rm c}+1)$ and $N_{\rm c}$, respectively.  More generally, we can replace $N_{\rm c}$ by an effective occupation number $N_{\rm eff}$, which, similar as in the cooling discussion, also takes into account of the finite overlap of the spectral peaks. Under this assumption we obtain a `thermal' two-mode squeezing master equation of the form
\begin{equation}\label{eq:ThermalTMS_ME}
\begin{split}
\mathcal{L}_{\rm sq} \hat{\rho} \simeq &\frac{\Gamma^a_{\rm sq}}{2} (N^a_{\rm eff}+1) \mathcal{D}[\hat A] + \frac{\Gamma^a_{\rm sq}}{2} N^a_{\rm eff} \mathcal{D}[\hat A^\dag]\\
+&\frac{\Gamma^b_{\rm sq}}{2}(N^b_{\rm eff}+1) \mathcal{D}[\hat B] + \frac{\Gamma^b_{\rm sq}}{2} N^b_{\rm eff} \mathcal{D}[\hat B^\dag].
\end{split}
\end{equation}
The rates $\Gamma^{a,b}_{\rm sq}$ and the effective occupation numbers $N^{a,b}_{\rm eff}$ can be estimated from the fluctuation spectrum given in Eq.~\eqref{Eq:SFFdef} as
\begin{equation}
\Gamma^{\eta=a,b}_{\rm sq} =\frac{S_{\rm FF}(\omega_\eta)-S_{\rm FF}(-\omega_\eta)}{2\cosh(r)},
\end{equation}
and
\begin{equation}
N^{\eta=a,b}_{\rm eff} =\frac{S_{\rm FF}(-\omega_\eta)}{S_{\rm FF}(\omega_\eta)-S_{\rm FF}(-\omega_\eta)} \geq N_{\rm c}.
\end{equation}
Then Eq.~\eqref{eq:ThermalTMS_ME} becomes exact in the sideband-resolved regime $\omega_{a,b}\gg \Gamma_{\rm ph}$ and reproduces as well the correct cooling dynamics in the limit $r\rightarrow 0$. While under general conditions Eq.~\eqref{eq:ThermalTMS_ME} is only a crude approximation, it is still expected to give accurate predictions for the squeezing parameter in the relevant regime $N_{\rm eff}^{a/b}<1$.  

\begin{figure}[]
  \centering
    \includegraphics[width=0.48\textwidth]{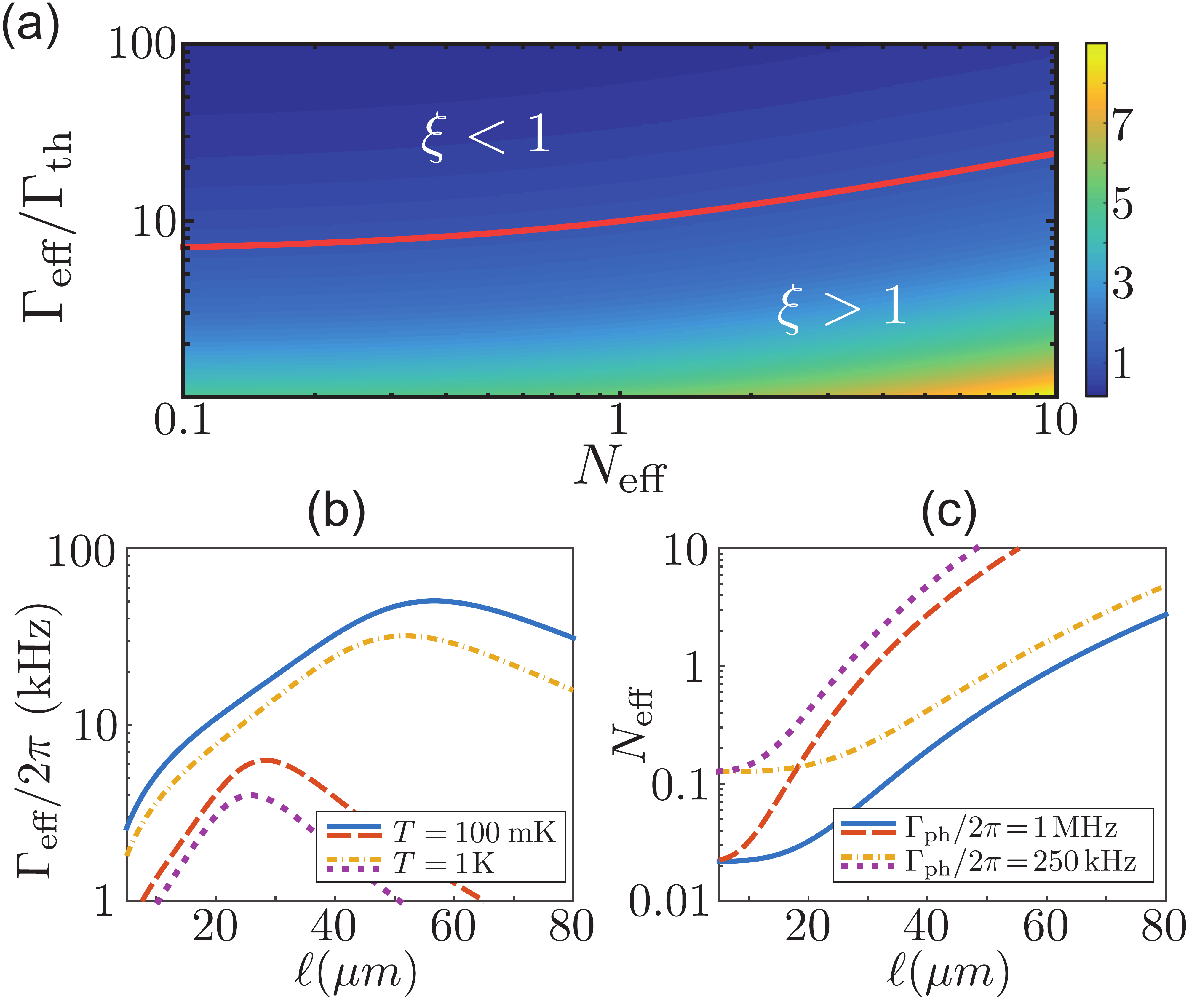}
      \caption{Contour plot of the minimal achievable separability parameter $\xi_{\rm min}={\rm min}\{\xi(r)|r\}$ as a function of the ratio $\Gamma_{\text{eff}} / \Gamma_{\text{th}}$ and the effective occupation $N_{\text{eff}}$. The red line corresponds to the threshold given by the condition $\xi_{\rm min} = 1$. For this plot the parameter $\xi(r)$ has been evaluated using Eq.~\eqref{eq:xitherm}, where equal parameters for the two modes and
$\gamma_a=\gamma_b \ll \Gamma_{\rm sq}$ have been assumed. (b) The values of $\Gamma_{\text{eff}}$ and $N_{\text{eff}}$ are plotted as a function of the length $\ell$ of the beam. The values for the support temperature $T$ and $\Gamma_{\text{ph}}$ for each line are given in the insets. All the other parameters are the same as in Fig.~\ref{Fig:nFinal}.}\label{SeparabilityContourFig}
\end{figure}

In the simplest case of identical mechanical modes, i.e.~$\gamma_aN_a= \gamma_b N_b=\Gamma_{\rm th}$, $N^{a}_{\rm eff}\simeq N^b_{\rm eff} \simeq N_{\rm eff}$, $\Gamma^{\eta=a,b}_{\rm sq}=\Gamma_{\rm eff}/(2\cosh r)$, and assuming again that $\gamma_{a,b}\ll \Gamma_{\rm sq}$, we can derive from Eq.~\eqref{eq:ThermalTMS_ME} a generalized expression for the separability parameter,
%
%In Fig.~\ref{SeparabilityContourFig} we plot the separability parameter $\xi$ against the ratio $\Gamma_{\text{eff}} / \Gamma_{\text{th}}$ and the effective occupation $N_{\text{eff}}$. The evaluation is done in the limit $\gamma_{a,b}\ll \Gamma_{\rm sq}$ and  $\gamma_aN_a= \gamma_b N_b=\Gamma_{\rm th}$. In this limit, the separability parameter is given by
\begin{equation}\label{eq:xitherm}
\xi \approx  \frac{4\Gamma_{\rm th}}{\Gamma_{\rm eff}} \cosh(r) + (1 + 2 N_{\rm eff}) e^{-2r},
\end{equation}
which reproduces Eq.~\eqref{Eq:xiApprox} for $N_{\text{eff}} = 0$. In Fig.~\ref{SeparabilityContourFig} (a) we plot the value of $\xi$ minimized with respect to $r$ for different ratios $\Gamma_{\text{eff}} / \Gamma_{\text{th}}$ and effective reservoir occupation numbers $N_{\text{eff}}$. This plot provides a general overview on the minimal conditions required for the dissipative preparation of entangled mechanical states. In particular,  it shows that the two-mode squeezing scheme does not rely on a strict zero reservoir temperature. Indeed, the steady-state entanglement is rather robust with respect to $N_{\rm eff}$, which is related to the fact that the thermal fluctuations of the environment become squeezed as well. However, note that for $N^{a/b}_{\rm eff}\gtrsim1$ the current approach can only provide a crude approximation for the squeezing parameter.

For the current setup, the expected values for $\Gamma_{\rm eff}$ and $N_{\rm eff}$ are plotted in Figs.~\ref{SeparabilityContourFig} (b) and (c) as a function of the beam length and for different values of $\Gamma_{\rm ph}$ and the support temperature $T$. For $T=100$ mK and $Q=10^6$ the relevant thermalization rate is $\Gamma_{\rm th}/2\pi \approx 2$ kHz, which shows that for $\ell\approx 30-70\,\mu$m the conditions for steady-state entanglement can be reached. Importantly, since the fundamental limit, $N_{\rm eff}\simeq N_{\rm c}$, remains small for temperatures up to $T\approx 4$K, a slightly improved coupling constant or Q-value would enable the dissipative generation of entanglement at these more convenient temperatures.

%The behaviour of $\Gamma_{\text{eff}}$ and $N_{\text{eff}}$, by varying the length $\ell$ of the beam, is given in Fig.~\ref{GeffNeffFig}.

\section{Conclusions}\label{sec:Conclusions}
In summary we have investigated a new approach for realizing mechanical ground-state cooling and dissipative state preparation schemes for mechanical beams, which makes use of the intrinsic reservoir of high-frequency compression modes. For the relevant example of a diamond nanobeam, we have shown that a single \SiV center can be used to engineer an effective mechanical reservoir, which can serve as a general tool for the preparation of ground-, squeezed- and entanglement states of low frequency vibrational modes. The use of a phononic bath can be advantageous, whenever no efficient coupling to optical dissipation channels is available, or when optical heating is a problem. Compared to cooling and reservoir engineering schemes with superconducting qubits, the large ground state splitting of the \SiV defect leads to higher cooling ratios and the possibility to achieve ground state cooling and entanglement already at experimentally convenient temperatures of  $T\sim 1$K. 

Beyond its potential use for mechanical state preparation, the current analysis also illustrates the rather strong magnetic and strain coupling of the \SiV ground state to both discrete and continuous phononic degrees of freedom. The use of such interactions for mechanical spin manipulations or phonon mediated couplings of distant centers is subject of current investigations~\cite{unpublished} and could open up a whole new range of mechanical control schemes for spin qubits in diamond.

%%%%%%%%%%%%%
\acknowledgements
We thank S. Meesala, A. Sipahigil, M. Loncar, A. Bleszynski Jayich and C. Becher for stimulating discussions. This work was supported by the Austrian Science Fund (FWF) through SFB FOQUS F40 and the START grant Y 591-N16. J.R.M acknowledges support from Fondecyt-Conicyt grant No 1141185.

%%%%%%%%%%%%%%%%%%%%%%%%%%%%%%%%%%%%%%%%%%%%%%%%%%%%%%%%%%%%%%%%%%%%%%%%%%%%%%%%%%%%%%%%%%%%

\appendix

\section{Effects of the JT interaction} \label{App:JTint}

In this section we present in more details the effects of the dynamical JT interaction~\cite{AbtewPRL2011,SlonczewskiPR1963} on the electronic structure of the \SiV defect in diamond. More precisely, we show that the JT interaction leads to a distortion of the orbital eigenstates of the dominant spin-orbit coupling. Such a perturbation has direct consequences on the strain-induced coupling to the high frequency phonon modes of the host crystal.

%We first recall the Hamiltonian describing the level splitting within the ground-state of a \SiV defect in an external magnetic field along its symmetry axis $\vec{B} = B_0 \vec{e}_z$ [cf.~Eq.~\eqref{Eq:HSiV}]~\cite{Hepp_2014}, 
%\begin{align}
%	\hat{H}_{\rm SiV} = -\lambda_\textrm{SO}\hat{L}_z\hat{S}_z + \hat{H}_\textrm{JT} + f\gamma_L B_0 \hat{L}_z  + \gamma_S B_0\hat{S}_z. 
%	\label{Eq:HSiVApp}
%\end{align}

In the basis spawned by the degenerate eigenstates $\ket{e_x, \uparrow}, \ket{e_x, \downarrow}, \ket{e_y, \uparrow}$ and $\ket{e_y,\downarrow }$, the different contributions to the \SiV energy levels, introduced in Eq.~\eqref{Eq:HSiV}, read
\begin{align}
\begin{split}
	(\gamma_SB_0-\lambda_{SO}\hat{L}_z)\hat{S}_z & =
	\frac{1}{2}\begin{bmatrix}
	\gamma_SB_0 & i\lambda_{SO} \\
	-i\lambda_{SO} & \gamma_SB_0
	\end{bmatrix} 
	\otimes
	\begin{bmatrix}
	1 & 0 \\
	0 & -1
	\end{bmatrix}, \\
	\hat{H}_{JT} & = \begin{bmatrix}
	\Upsilon_x & \Upsilon_y  \\
	\Upsilon_y & -\Upsilon_x
	\end{bmatrix} \!\otimes\! 
	\begin{bmatrix}
	1 & 0 \\
	0 & 1
	\end{bmatrix}.
\end{split}
\label{Eq:HSiVApp}
\end{align}
Here, $\Upsilon_x$ ($\Upsilon_y$) represents an energy shift due to local strain in the crystal along $x$ ($y$), such that $\Upsilon=\sqrt{\Upsilon_x^2+\Upsilon_y^2}$. As in the main text, we neglect the effect of the reduced orbital Zeeman interaction ($\sim f\gamma_L B_0$). Diagonalizing Eq.~\eqref{Eq:HSiVApp} leads to the eigenstates 
\begin{align}
\begin{split}
        \vert 1 \rangle & = \left(\cos\theta \vert e_x \rangle - i  \sin\theta e^{-i\phi} \vert e_y \rangle\right)\vert\!\downarrow \rangle, \\
	\vert 2 \rangle & = \left(\cos\theta \vert e_x \rangle + i\sin\theta e^{i\phi}\vert e_y \rangle\right)\vert\!\uparrow \rangle, \\
	\vert 3 \rangle & = \left(\sin\theta \vert e_x \rangle + i  \cos\theta e^{-i\phi}\vert e_y \rangle\right)\vert\!\downarrow \rangle, \\
	\vert 4 \rangle & = \left(\sin\theta \vert e_x \rangle - i  \cos\theta e^{i\phi}\vert e_y \rangle\right)\vert\!\uparrow \rangle,
\end{split}
\label{Eq:StatesApp}
\end{align} 
where
%\begin{align}
%	\tan(\theta) = \frac{\sqrt{\lambda_{\rm SO}^2 + 4\Upsilon_y^2}
%}{\vert\Delta -2\Upsilon_x\vert}, \qquad
%	\tan(\phi) = \frac{\lambda_{\rm SO}}{2\Upsilon_y}.
%\end{align}
\begin{align}
	\tan(\theta) = \frac{2\Upsilon_x + \Delta}{\sqrt{\lambda_{\rm SO}^2 + 4\Upsilon_y^2}}, \qquad
	\tan(\phi) = \frac{2\Upsilon_y}{\lambda_{\rm SO}}.
\end{align}
The corresponding eigenenergies are:
\begin{align}
	E_{3,1} = (-\gamma_SB_0 \pm \Delta)/2, \quad E_{4,2} = (\gamma_SB_0 \pm \Delta)/2.
	\label{Eq:EnergApp}
\end{align}

One of the consequences of the orbital mixing by the JT interaction that is relevant for the present scheme is the possibility of new strain-mediated processes within the ground state. 
While $\hat H_{\rm strain}$ [Eq.~\eqref{Eq:Hstrain1}] can only mediate the transitions $\ket{1} \leftrightarrow \ket{3}$ and $\ket{2} \leftrightarrow \ket{4}$ in absence of the JT effect, orbital distortions due to the JT effects leads to
\begin{align}
	\hat{H}^{\rm JT}_\textrm{strain} & = 2g_1\cos\theta\sin\theta (\hat{\gamma}_{xx} - \hat{\gamma}_{yy}) (\ket{1}\bra{3} + \ket{2}\bra{4}) \nonumber \\
	& - 2ig_2(\sin^2\theta e^{i\phi} + \cos^2\theta e^{-i\phi}) \hat{\gamma}_{xy} (\ket{1}\bra{3} + \ket{4}\bra{2}) \nonumber \\
	& + \left[\frac{g_1}{2}(\cos^2\theta - \sin^2\theta)(\hat{\gamma}_{xx} - \hat{\gamma}_{yy}) \right. \nonumber \\
	& \qquad  \qquad \quad + \left. 2g_2\cos\theta\sin\theta\sin\phi \hat{\gamma}_{xy} \right] \nonumber \\
	& \qquad (\ket{1}\bra{1}+\ket{2}\bra{2}-\ket{3}\bra{3}-\ket{4}\bra{4}) + \rm{H.c.}
\end{align}
For realistic JT interaction strengths~\cite{HeppPRL2014}, the coupling constant of the last term can become comparable to the others. However, those processes will always be off-resonant for the schemes presented in this work, and its effects negligible. 
In other contexts, e.g.~where long-lived coherent superposition of electronic states are necessary, the distortion due to the JT could add to decoherence and thus might be important to consider.

%%%%%%%

\section{Strain coupling Hamiltonian} \label{App:HStrain}

Local strain or local distortions in the \SiV structure generates a collective displacement of the defect atoms. This leads to a change in the potential seen by each atoms and result in a modification of the electronic distribution of the defect via electron-ion interaction. To first order in the ion displacements and in the Born-Oppenheimer approximation, this local distortion effect can be modeled by the strain Hamiltonian 
\begin{equation}
\hat{H}_{\mbox{\scriptsize strain}} = \sum_{i,j,\alpha,\beta}|\alpha \rangle \langle \alpha |\hat V_{ij} | \beta \rangle\langle \beta | \hat{\gamma}_{ij}.
\end{equation}
Here, $|\alpha\rangle$ is the electronic basis and $\hat V_{ij}$ are couplings that involve the electron-ion Coulomb interaction~\cite{MazeNJP2011}. The strain field tensor can be symmetrically decomposed as $\hat{\gamma} = \hat{\gamma}_{A_{1g}} + \hat{\gamma}_{E_g}$, where
\begin{align}
\begin{split}
\hat{\gamma}_{A_{1g}} & = \left[\begin{array}{ccc} 
                                      {1 \over 2}\left(\hat{\gamma}_{xx} + \hat{\gamma}_{yy} \right) & 0 & 0 \\
                                     0 & {1 \over 2}\left(\hat{\gamma}_{xx} + \hat{\gamma}_{yy} \right) & 0 \\
                                     0 & 0 & \hat{\gamma}_{zz} \\
                                      \end{array} \right], \\
\hat{\gamma}_{E_{g}} & = \left[\begin{array}{ccc} 
                                      {1 \over 2}\left(\hat{\gamma}_{xx} - \hat{\gamma}_{yy} \right) & \hat{\gamma}_{xy} & \hat{\gamma}_{xz} \\
                                     \hat{\gamma}_{xy} & {1 \over 2}\left(\hat{\gamma}_{yy} - \hat{\gamma}_{xx} \right) & \hat{\gamma}_{yz} \\
                                     \hat{\gamma}_{xz} & \hat{\gamma}_{yz} & 0 \\
                                       \end{array} \right].
\end{split}
\end{align}
Due to the inversion symmetry of the SiV$^-$~\cite{GossPRL1996}, the orbital degrees of freedom of the states within the ground and excited subspace are characterized by parity~\cite{TinkhamBook}. As a consequence, expectation values 
$\langle \alpha | \hat V_{iz} | \alpha \rangle$ and $\langle \alpha | \hat V_{zz} | \alpha \rangle$ vanish in both ground and excited subspaces. Therefore, in the electronic basis spawned 
by $\{|e_{gx} \rangle, |e_{gy} \rangle \}$, the strain Hamiltonian can be written as~\cite{HeppPhD}
\begin{equation}
\hat{H}_{\mbox{\footnotesize strain}} = \left[ \begin{array}{cc} 
                                                                                    \delta & 0 \\
                                                                                    0 & \delta 
                                                                                \end{array} \right] +  \left[ \begin{array}{cc} 
                                                                                    \alpha & \beta \\
                                                                                    \beta & -\alpha 
                                                                                \end{array} \right],
\end{equation}
with $\delta =   g_0  \left(\hat{\gamma}_{xx} + \hat{\gamma}_{yy} \right)$,  
$\alpha =   g_1  \left(\hat{\gamma}_{xx} - \hat{\gamma}_{yy} \right)$ and $\beta =  g_2  \hat{\gamma}_{xy}$. Here, 
$g_0,g_1$ and $g_2$ are coupling constants. The first term of the strain Hamiltonian is the energy shift induced 
by symmetry local distortions and can be neglected. 
Finally, if we write the strain Hamiltonian using the basis spawned by the eigenstates of the spin-orbit coupling [see after Eq.~\eqref{Eq:HSiV}], we recover Eq.~\eqref{Eq:Hstrain1}, i.e. 
\begin{align}
\hat{H}_{\mbox{\footnotesize strain}} & =  \left[\begin{array}{cc} 
                                                                                    0 & \alpha -i \beta \\
                                                                                    \alpha + i \beta & 0 
                                                                                \end{array} \right] \otimes  \left[ \begin{array}{cc} 
                                                                                    1 & 0\\
                                                                                   0 &  1
                                                                                \end{array} \right]  \\ 
& = g_1 \left(\hat{\gamma}_{xx} - \hat{\gamma}_{yy}  \right) \left(\hat{L}_{-}+\hat{L}_{+} \right)  - ig_2 \hat{\gamma}_{xy}\left(\hat{L}_{-}-\hat{L}_{+} \right). \nonumber
\end{align}

%%%%%%%%%%%%%%%%%%%%%%%%%%%%%%%%%%%%%%%%%%%%%%%%%%%%%

\section{Phonon spectral density in the 3D limit} \label{App:J3D}

In this appendix we derive in more details the spectral density of the high-frequency phonon modes in the limit where the \SiV defect is embedded in an infinite bulk diamond structure. In contrast to the strict 1D limit, where only longitudinal modes propagate along the beam, the 3D structure can host longitudinal and transverse phononic waves in every directions. Consequently, for every propagation directions $\vec{q}$, there are three possible polarizations with the corresponding displacements 
\begin{equation}
\begin{split}
	\hat{\mathbf{u}}_l(\vec{r},t) & = \sum_{\vec{q}} \frac{1}{\sqrt{2\rho V \omega_{l,q}}}\vec{e}_q\left[ \hat{c}^\dag_{l,\vec{q}}(t)e^{-i\vec{q}\cdot\vec{r}} + \hat{c}_{l,\vec{q}}(t)e^{i\vec{q}\cdot\vec{r}}\right], \\
	\hat{\mathbf{u}}_\theta(\vec{r},t) & = -\sum_{\vec{q}} \frac{1}{\sqrt{2\rho V \omega_{t,q}}}\vec{e}_\theta\left[ \hat{c}^\dag_{\theta,\vec{q}}(t)e^{-i\vec{q}\cdot\vec{r}} + \hat{c}_{\theta,\vec{q}}(t)e^{i\vec{q}\cdot\vec{r}}\right], \\
	\hat{\mathbf{u}}_\phi(\vec{r},t) & = \sum_{\vec{q}} \frac{1}{\sqrt{2\rho V \omega_{t,q}}}\vec{e}_\phi\left[ \hat{c}^\dag_{\phi,\vec{q}}(t)e^{-i\vec{q}\cdot\vec{r}} + \hat{c}_{\phi,\vec{q}}(t)e^{i\vec{q}\cdot\vec{r}}\right].
\end{split}
\label{Eq:3DModes}
\end{equation}
Here, $l$ stands for the longitudinal mode, with a displacement along the unit vector $\vec{e}_q$, $\theta$ and $\phi$ denote the two transverse modes, with displacements along the unit vectors $\vec{e}_\theta$ and $\vec{e}_\phi$ respectively. The operators $\hat{c}_{i,\vec{q}}$ are bosonic lowering operator for the mode with wavevector $\vec{q}$ and polarization $i$, and $V$ is the volume of the diamond structure. Neglecting the weak Poisson ratio of diamond, the mode frequencies are given by
\begin{align}
	\omega_{l,q} = v_{l} q \approx \sqrt{\frac{E}{\rho}}q, 
	\qquad
	\omega_{t,q} = v_{t} q = \sqrt{\frac{E}{2\rho}}q,
\end{align}
where $v_l$ and $v_t$ are  sound velocities of the longitudinal and transverse modes, respectively. 

It is straightforward to compute the strain fields $\hat{\gamma}_{ij} = (\partial\hat{u}_i/\partial x_j + \partial\hat{u}_j/\partial x_i)/2$ at the \SiV position for each of the phonon modes of Eqs.~\eqref{Eq:3DModes} and substitute them in Eq.~\eqref{Eq:Hstrain}, leading to the strain interaction Hamiltonian $\hat{H}^{3D}_{\rm strain} = \hat{H}^{l}_{\rm strain} + \hat{H}^{\theta}_{\rm strain} + \hat{H}^{\phi}_{\rm strain}$. As an example, the contribution from the transverse modes $\hat{c}_{\theta,\vec{q}}$ reads:
\begin{equation}
\begin{split}
	\hat{H}_\textrm{strain}^\theta  & = \sum_{\vec{q}} ig_{1,\theta} (\vec{q})\hat{J}_+ \hat{c}_{\theta,\vec{q}} - g_{2,\theta}(\vec{q})\hat{J}_+ \hat{c}_{\theta,\vec{q}} + \rm{H.c.},  \\
	 g_{1,\theta}(\vec{q}) & = \frac{g_1}{v_t}\sqrt{\frac{\omega_{t,q}}{2\rho V }}\sin\theta\cos\theta(\sin^2 \phi - \cos^2 \phi), \\
	g_{2,\theta}(\vec{q}) & = \frac{2g_2}{v_t}\sqrt{\frac{\omega_{t,q}}{2\rho V }}\sin\theta\cos\theta\sin\phi\cos\phi.
\end{split}
\end{equation}
%\begin{align}
%	\hat{H}_\textrm{strain}^\theta  & = -\frac{i}{c_t}\sum_{\vec{q}}\sqrt{\frac{\omega_{t,q}}{2\rho V }}\sin\theta\cos\theta\left[\lambda_1(\sin^2 \phi - \cos^2 \phi)(\hat{J}_- + \hat{J}_+) \right. \nonumber \\
%	& \left. -2i\lambda_2\sin\phi\cos\phi (\hat{J}_- - \hat{J}_+)\right]\left( \hat{c}^\dag_{\theta,\vec{q}} - \hat{c}_{\theta,\vec{q}}\right),
%\end{align}
%with $\hat{H}^{3D}_{\rm strain} = \hat{H}^{l}_{\rm strain} + \hat{H}^{\theta}_{\rm strain} + \hat{H}^{\phi}_{\rm strain}$. 

The spectral density, $J^{3D}(\omega) = J^{l}(\omega) + J^{\theta}(\omega) + J^{\phi}(\omega)$, is obtained by substituting the different coupling constants $g(\vec{q})$ in Eq.~\eqref{Eq:SpecFunc},
\begin{align}
	J^{\theta}(\omega) = \sum_{\vec{q}} \left[ \vert g_{1,\theta}(\vec{q}) \vert^2 + \vert g_{2,\theta}(\vec{q}) \vert^2 \right] \delta(\omega - \omega_{t,q}),
\end{align}
leading to the infinite 3D limit
\begin{align}
	J^{3D}(\omega) & = \frac{g_1^2 + g_2^2}{\pi\rho}\left( \frac{1}{5v_t^5} + \frac{2}{15v_l^5} \right)\omega^3.
\end{align}
In contrast to the 1D limit, the spectral density does not depend on the position of the \SiV defect and scales as $\omega^3$ (instead of $\omega$) due to the higher density of states at high frequencies. Under the approximation $g_1=g_2$ and $\sqrt{2}v_t=v_l=v$ we obtain the result stated in Eq.~\eqref{Eq:J3D}.

%\bibliographystyle{apsrev}
%\bibliography{MALPapersRefs}

%Two-mode squeezing Nori  	Phys. Rev. B 76, 064305 (2007)
\end{document}